\documentclass[12pt]{article}
\usepackage{amssymb,amsmath,amsfonts,eurosym,geometry,ulem,graphicx,caption,color,setspace,sectsty,comment,footmisc,caption,pdflscape,subcaption,dsfont,natbib,hyperref
}
\usepackage{multirow}
\usepackage{array,sgame,istgame }
\usepackage{fouriernc}
\usepackage[flushleft]{threeparttable}
\usepackage{nicematrix}

\usepackage{algorithm,algpseudocodex} %display algorithm
\usepackage{adjustbox}

\newcolumntype{L}[1]{>{\raggedright\let\newline\\arraybackslash\hspace{0pt}}m{#1}}
\newcolumntype{C}[1]{>{\centering\let\newline\\arraybackslash\hspace{0pt}}m{#1}}
\newcolumntype{R}[1]{>{\raggedleft\let\newline\\arraybackslash\hspace{0pt}}m{#1}}

\begin{document}

\begin{titlepage}
\title{
\fontsize{18pt}{18pt}\selectfont
Algorithmic Collusion under Observed Demand Shocks
}

\author{Zexin Ye
\thanks{ye.754@osu.edu, Department of Economics, The Ohio State University.
I would like to thank Huanxing Yang for his valuable advice. 
I am also grateful to Yonghong An, Yaron Azrieli, Paul J. Healy, and Lixin Ye for their insightful comments. 
I thank Bill Wang and Chengcheng Wang for helpful discussions, and I also appreciate the worthwhile feedback from the OSU Theory/Experimental Reading Group and the 2025 Stony Brook International Conference on Game Theory.
All errors remain my own.
}}
\date{\today}
\maketitle

\begin{abstract}
\noindent
% motivation
This paper examines how the observability of demand shocks influences pricing patterns and market outcomes when firms delegate pricing decisions to Q-learning algorithms.
% result
Simulations show that demand observability induces Q-learning agents to adapt prices to demand fluctuations, giving rise to distinctive demand-contingent pricing patterns across the discount factor $\delta$, consistent with \citet{rs1986}. When $\delta$ is high, they learn procyclical pricing, charging higher prices in higher demand states. In contrast, at low $\delta$, they lower prices during booms and raise them during downturns, exhibiting countercyclical pricing. Q-learning agents also autonomously sustain supracompetitive profits, indicating that demand observability does not hinder algorithmic collusion. I further explore how the information available to algorithms shapes their learned pricing behavior.
% contribution
Overall, the results suggest that, through pure trial and error, Q-learning algorithms internalize both the stronger deviation incentives during booms and the trade-off between short-term gains and long-term continuation values governed by the discount factor, thereby reproducing the cyclicality of pricing patterns predicted by collusion theory.
\\
\vspace{0in}
\vspace{0in}

% D21	Firm Behavior: Theory
% D43	Oligopoly and Other Forms of Market Imperfection
% D83	Search • Learning • Information and Knowledge
% L13	Oligopoly and Other Imperfect Markets
        
\bigskip
\end{abstract}
\setcounter{page}{0}
\thispagestyle{empty}
\end{titlepage}
\pagebreak \newpage

\doublespacing

\section{Introduction} \label{sec:introduction}
%%%%%%%%%%%%%%%%% background: algorithmic collusion %%%%%%%%%%%%%%%%%
% motivation
In recent years, AI-powered pricing algorithms have been widely adopted across industries, reshaping how firms compete and set prices. While these algorithms offer clear convenience, they have also raised growing concerns among researchers and policymakers regarding their effects on competition and consumer welfare (\citeauthor{calvano2020protecting}, \citeyear{calvano2020protecting}; see \citeauthor{review2021ac}, \citeyear{review2021ac} for a comprehensive review).
A notable example comes from the U.S. rental housing market, where many landlords have adopted commercial pricing software developed by RealPage to help them adjust rents in response to local market conditions.\footnote{RealPage collects proprietary data such as apartment prices and occupancy rates from subscribing landlords and uses this information to estimate market demand and dynamically recommend rental rates.} 
The widespread use of this software has been seen as contributing to higher rental prices, affecting millions of households and adding an estimated $\$3.8$ billion to annual rental expenditures. In response, a lawsuit has been filed against RealPage and, more broadly, a Senate bill has been introduced to address algorithmic collusion.\footnote{For the lawsuit, see https://www.justice.gov/archives/opa/media/1364976/dl?inline. For the Senate bill, see https://www.congress.gov/bill/118th-congress/senate-bill/3692.}\footnote{\citet{abada2025algorithmic} give an inclusive definition of algorithmic collusion:
“Algorithmic collusion is when supracompetitive outcomes are produced by learning algorithms without human design to produce those outcomes.”}

This example, together with dynamic pricing systems used by airlines, hotels, and platforms such as Amazon, Airbnb, and Uber, illustrates a key feature of such algorithms: their ability to adjust prices in response to changing market conditions \citep{elmaghraby2003dynamic, cohen2016using, calder2023coordinated, spann2025algorithmic}.
This naturally raises a question: how do market fluctuations, such as stochastic demand shocks, shape the learning outcomes of pricing algorithms, especially when these algorithms learn autonomously?

% gap
Existing studies have shown that autonomous learning algorithms, through repeated interaction, can learn to sustain supracompetitive outcomes without explicit programming or human design \citep{calvano2020, klein2021, calvano2021, asker2024impact}. However, most of this work focuses on environments with fixed demand or unobserved demand shocks, and the implications of settings in which algorithms can respond to demand fluctuations remain largely unexplored.
The effect of demand observability is ambiguous. On one hand, more predictable demand reduces uncertainty and enables algorithms to learn more precise mappings from demand states to optimal prices, thereby facilitating collusive coordination. On the other hand, richer demand information may intensify deviation incentives during booms, as the short-term gains from undercutting can outweigh the expected future losses from punishment, thereby undermining the sustainability of collusion.

% setting
Motivated by these policy concerns and the existing research gap, this paper examines algorithmic pricing in a repeated setting with stochastic demand shocks, where firms delegate pricing to algorithms that set prices after observing the current demand state. This setup allows me to adopt the framework of \citet{rs1986}.\footnote{Following \citet{rs1986}, I assume that the current demand shock is publicly observable. This assumption can be interpreted as firms either directly observing realized demand or having access to perfect demand forecasts.}
In the model, two agents engage in an infinitely repeated Bertrand competition game with a homogeneous product. In each period, an i.i.d. demand shock occurs, and after observing the realized demand state, both agents simultaneously choose prices. Past prices are publicly observed.

Both agents are implemented with Q-learning algorithms, as commonly adopted in the literature of algorithmic pricing. Q-learning is a fundamental reinforcement learning method valued for its simplicity and model-free design.\footnote{Reinforcement learning (RL) has become a foundational technique in modern artificial intelligence. See \citet{sutton2018reinforcement} for a comprehensive introduction.} I incorporate the realized demand state into the state representation, allowing Q-learning agents to condition their pricing decisions on observed demand shocks.\footnote{Technical details are provided in Section~\ref{sec:design}.}

% finding 
My main findings are threefold. 
First, Q-learning agents adapt their pricing to demand fluctuations, learning distinct pricing patterns as the discount factor $\delta$ changes, consistent with \citet{rs1986}. 
When $\delta$ is high, they coordinate on procyclical pricing, charging higher prices in higher demand states. Conversely, when $\delta$ is low, they switch to countercyclical pricing: greater weight on immediate rewards strengthens deviation incentives in booms, making high prices unsustainable there but sustainable in downturns. Consequently, agents lower prices in booms while maintaining relatively high prices in downturns.\footnote{As $\delta$ decreases further, even supracompetitive prices in downturns become unsustainable, leading to rigid pricing at the competitive level across demand states.}

Second, Q-learning agents can autonomously learn to sustain supracompetitive profits under observed demand shocks, indicating that demand observability does not hinder algorithmic collusion. However, relative to the fixed-demand benchmark, learning under observed demand shocks exhibits a profit reversal: it delivers higher profits when $\delta$ is low but lower profits when $\delta$ is high.\footnote{The fixed-demand benchmark is constructed by training Q-learning agents separately in each fixed demand state and averaging profits weighted by the demand distribution.} This profit reversal arises from the demand-contingent pricing, highlighting that the observability of demand shocks is a double-edged sword for profitability.

Last, I examine how memory, specifically past demand realizations and past price pairs, shapes demand-contingent pricing. Removing past demand makes it difficult for algorithms to distinguish whether a previous price change reflects a demand shock or a rival’s deviation, thereby weakening demand-contingent pricing and shifting learning toward rigid pricing with identical prices across demand states.
The effect of price memory, however, depends on the discount factor. When $\delta$ is high, Q-learning agents can still sustain procyclical pricing without recalling past prices. When $\delta$ is low, by contrast, the absence of price memory eliminates countercyclical pricing entirely, with all simulations invariably converging to rigid pricing at the competitive level.

% contribution
Overall, this paper bridges classic collusion theory and modern reinforcement learning by showing that Q-learning algorithms, when exposed to observed demand shocks, experience payoff variation across demand states and prices, thereby autonomously reproducing the cyclicality of pricing patterns predicted by \cite{rs1986}. These learning outcomes suggest that, despite having no explicit understanding of equilibrium or intertemporal incentives, the algorithms, through pure trial and error, come to internalize the key insights underlying this cyclicality, namely that (i) stronger deviation incentives arise during booms and (ii) the trade-off between short-term gains and long-term continuation values is governed by $\delta$, and behave as if guided by them.

The rest of the paper is organized as follows. Section \ref{sec:literature} reviews the related literature. Section \ref{sec:design} presents the economic environment for the simulations and explains the modified Q-learning algorithms that observe demand shocks. Section \ref{sec:pattern} then describes how price cycles are derived, based on which I analyze demand-contingent pricing patterns and evaluate their performance. Section \ref{sec:state_outcome} investigates how memory shapes demand-contingent pricing. Section \ref{sec:robust} conducts a series of robustness checks. Section \ref{sec:conclusion} concludes.
%%%%%%%%%%%%%%%%%%%%%%%%%%%%%%%%%%%%%%%%%%%%%%%%%%%%%%%%%%%%%%%%%%%%%%%%%%%%%

%%%%%%%%%%%%%%%%%%%%%%%%%%%%%%%%%%%%%%%%%%%%%%%%%%%%%%%%%%%%%%%%%%%%%%%%%%%%%
\section{Literature Review} \label{sec:literature}
This paper relates to three strands of literature.
The first is the rapidly growing body of research showing that pricing algorithms can autonomously learn to achieve collusive outcomes. Early work dates back to \citet{waltman2008q}. More recently, \citet{calvano2020} reignite this field by investigating how Q-learning algorithms interact in a repeated Bertrand competition game with logit demand. They show that Q-learning agents learn to set supracompetitive prices, sustained by collusive strategies that effectively deter deviations.
\citet{klein2021} applies Q-learning algorithms to the dynamic sequential-pricing framework of \citet{mt1988} and finds that they can still learn to sustain stable collusion. Building on the framework of \citet{gp1984}, in which firms face unobserved stochastic demand shocks and cannot perfectly monitor their rivals’ behavior, \citet{calvano2021} show that Q-learning agents are still able to sustain tacit collusion, confirming that imperfect monitoring does not preclude algorithmic collusion.\footnote{For other recent studies, \citet{fgs2024} show that agents powered by large language models can sustain autonomous collusion without explicit instructions. \citet{ballestero2021collusion} demonstrates that algorithmic collusion can also arise in sequential-pricing environments with stochastic costs.}
This line of research emphasizes that supracompetitive outcomes are supported by reward-punishment schemes learned through repeated interaction.\footnote{As defined by \cite{h2018ac}, a reward-punishment scheme rewards a firm for adhering to the supracompetitive outcome and punishes any deviation from it.}
However, recent studies show that a reward-punishment scheme is not a necessary condition for Q-learning agents to learn supracompetitive outcomes (see \citet{abada2025algorithmic} for a review).
For example, \citet{banchio2022artificial} show that collusive outcomes can emerge from spontaneous coupling between memoryless Q-learning agents.\footnote{Without memory, agents cannot monitor their rivals’ prices, making it impossible to detect or punish deviations.} \citet{lambin2024less} provide simulation evidence showing that the supracompetitive outcomes documented by \citet{calvano2020} and \citet{klein2021} can also be achieved in the memoryless setting.\footnote{\citet{asker2024impact} find that even when Q-learning agents are myopic, placing no value on the future, they can still converge to supracompetitive prices through asynchronous learning.}
Building on these studies, this paper investigates algorithmic pricing in a previously unexplored environment where demand shocks are observable, and examines how memory shapes learning outcomes.

Second, this paper contributes to an important strand of the collusion literature that studies how the demand environment affects firms’ ability to sustain collusion.
\citet{rs1986} study a setting in which demand shocks are modeled as i.i.d. and publicly observed. They show that when firms are sufficiently patient, they can sustain the best collusive outcome by charging the monopoly price in each demand state, leading to procyclical pricing, where prices rise during booms and fall during downturns. However, when the discount factor is lower, the incentive to deviate during booms becomes strong enough that firms must lower prices to mitigate the temptation to undercut.
As a result, unlike \citet{gp1984}, \citet{rs1986} predict that price wars occur during booms rather than downturns. In some cases, prices in booms may even fall below those in downturns,  resulting in countercyclical pricing, where prices move opposite to the business cycle.
Subsequent studies extend this framework by relaxing the i.i.d. assumption on demand shocks \citep{k1991correlated, hh1991cycle, bagwell1997collusion, knittel2010tacit}.
Another direction explores how the information structure of demand affects collusion.
\citet{mt2019demand} and \citet{ow2021demand} analyze how more precise public signals about the current demand state shape firms’ ability to coordinate on collusive pricing, while \citet{hy2007rigidity} consider a setting in which each firm receives a conditionally independent private signal about the underlying demand state.
\citet{sugaya2024collusion} further show that when the demand function is affine, the optimal information disclosure rule takes the form of upper censorship, under which firms receive a demand signal that pools high demand above a cutoff to stabilize collusion.
This paper examines whether Q-learning agents, without equilibrium reasoning and instead learning through trial and error, can endogenously reproduce the cyclicality of pricing patterns predicted by \citet{rs1986}.

In a broader context, this study contributes to the growing literature on algorithmic competition across diverse market environments.
For example, \citet{acex2024} provide empirical evidence that the widespread adoption of algorithmic pricing in Germany’s retail gasoline market significantly raises profit margins and thus softens competition. 
In the context of digital platforms, \citet{jrw2023} show that when sellers use pricing algorithms, platforms can implement demand-steering rules to destabilize algorithmic collusion, improving consumer welfare and increasing platform revenue.
In financial trading, \citet{dou2025ai} show that Q-learning agents can endogenously generate collusive outcomes through price-trigger strategies or learning biases.
\citet{bs2022} find that Q-learning agents learn to collude in first-price auctions but not in second-price auctions.
\cite{d2024} and \cite{xu2024mechanism} demonstrate that reinforcement learning agents can learn to cooperate or collude under certain conditions, depending on their learning rules and the structure of the game.
\citet{friedrich2024inventory} study algorithmic collusion in episodic markets with inventory constraints.

%%%%%%%%%%%%%%%%%%%%%%%%%%%%%%%%%%%%%%%%%%%%%%%%%%%%%%%%%%%%%%%%%%%%%%%%%%%%%

%%%%%%%%%%%%%%%%%%%%%%%%%%%%%%%%%%%%%%%%%%%%%%%%%%%%%%%%%%%%%%%%%%%%%%%%%%%%%
\section{Experimental Design} \label{sec:design}
\subsection{Economic Model}
Two firms engage in an infinitely repeated Bertrand competition game with a homogeneous product under linear demand.\footnote{Elimination of product differentiation facilitates a more tractable analysis of demand-contingent pricing.} 
In each period $t$, a random common demand shock $\theta_t$ realizes, shifting the market demand curve in parallel. This shock is i.i.d. and uniformly distributed on $[\underline{\theta},\overline{\theta}]$; 
After observing $\theta_t$, both firms simultaneously choose prices. 
Let $(p_{1t},p_{2t})$ denote the price pair and $p_{-it}$ the rival's price. Demand for firm $i$ in period $t$ is given by
$$
D_{it}(p_{i t}, p_{-i t}, \theta_t)= 
\begin{cases}

\theta_t-p_{i t} & \text { if } p_{i t}<p_{-it} \\ 

\dfrac{ \theta_t-p_{i t}}{2} & \text { if } p_{i t}=p_{-it}\\ 

0 & \text { if } p_{i t}>p_{-it}

\end{cases}
$$
\noindent Correspondingly, the period payoff for firm $i$ is 
$\pi_{it}=(p_{it}-c_i)  D_{it}(p_{i t}, p_{-i t}, \theta_t)$, where $c_i$ is the constant marginal cost.

\paragraph{Dynamic Problem}
In this repeated game, each firm seeks to maximize the expected present value of profits, $\mathbb{E}\left[ \sum_{t=0}^{\infty} \delta^t \pi_{it} \right]$, where $\delta \in (0,1)$ is the discount factor.
The problem can be formulated as a Markov decision process in which  firm $i$ chooses price $p_{it}$ based on state $s_t$. The value function for firm $i$ is
\begin{equation}
V_i(s) = \max_{p_i \in A} \Big\{
\pi_i + \delta \mathbb{E}\left[ V_i(s^{\prime}) \mid s, p_i\right]
\Big\} \label{valuefunc}
\end{equation}
where $s^{\prime}$ is the next state and $A$ is the action (price) space. 
% The specific details of the action space and state space in this model are discussed next.

\paragraph{Action}
As firms delegate their pricing decisions to Q-learning algorithms, I hereafter refer to them as Q-learning agents (or simply agents). Q-learning requires a finite action space, and thus the price set is discretized into $m$ evenly spaced points over $[\underline{p}^C, \,\overline{p}^M]$, where $\underline{p}^C$ and $\overline{p}^M$ denote the competitive (Bertrand Nash equilibrium) and monopoly prices under the lowest and highest demand states, respectively.

\paragraph{State}
The state $s_t$ consists of two components.
First, it includes the currently realized demand state $\theta_t$, allowing Q-learning agents to respond to demand fluctuations.
Second, under perfect public monitoring, it also includes the history of past price pairs and demand states, enabling agents to detect and punish deviations.
To prevent the state space from growing indefinitely, I follow \citet{calvano2020} and assume that Q-learning agents have bounded memory of length $K$.
Accordingly, the state variable is represented by the truncated history of length $K$ and the current demand state:
$$
s_t = \{ (p_{1t-K},p_{2t-K}, \theta_{t-K}), \ldots, (p_{1t-1},p_{2t-1},\theta_{t-1}),\theta_t \}
$$

\subsection{Q-Learning Algorithms}
The value function can be expressed in terms of a Q-function that represents the expected discounted value (Q-value) associated with selecting price $p$ in state $s$, as given by
\begin{equation}
Q_i(s,p) = 
\pi_i + \delta \mathbb{E}\left[ 
\max _{p^{\prime} \in A} Q_i(s^{\prime}, p^{\prime})
 \mid s, p\right]
 \label{Qfunc}
\end{equation}
where the first term on the right-hand side is the immediate reward and the second term
is the (conditional) expected continuation value.\footnote{The relationship between the Q-function and the value function is $V_i(s) \equiv \max\limits_{p \in A} Q_i(s, p)$.}
Since $S$ and $A$ are finite, the Q-function can be represented as an $|S| \times|A|$ matrix, in which each entry of the Q-matrix $\mathbf{Q}_{i}$ stores the corresponding Q-value $Q_{i}(s,p)$.

However, the Q-matrix $\mathbf{Q}_i$ cannot be solved analytically. Although each firm knows the demand distribution and the profit function, it does not know its rival’s pricing strategy. As a result, the realized reward $\pi_i(p_i, p_{-i}, \theta)$ is unpredictable ex ante, and the transition function $F_i(s^{\prime} \mid s, p_i)$ remains unknown.
To address this challenge, I employ the Q-learning algorithm of \citet{watkins1989}, a model-free reinforcement learning method that enables agents to approximate the optimal Q-function through trial-and-error interactions with the environment, without requiring prior knowledge of the transition process.\footnote{Although the demand distribution and the profit function are known to firms, Q-learning does not rely on this information when updating Q-values.} In this way, Q-learning agents iteratively update their Q-values using realized profits and observed state transitions.

\paragraph{Learning Equation}
In each period, agent $i$ updates the corresponding cell $(s_t, p_{i t})$ in $\mathbf{Q}_{it}$ according to
\begin{equation}
Q_{it+1}(s,p)=(1-\alpha)Q_{it}(s,p)+
\alpha\left[\pi_{it}+
\delta \max _{p^{\prime} \in A} Q_{it}(s^{\prime}, p^{\prime}) 
\right] \label{eq:learningfunc}
\end{equation}
where $ s^{\prime}=s_{t+1} = \left\{ (p_{1t-K+1}, p_{2t-K+1}, \theta_{t-K+1}), \ldots, (p_{1t},p_{2t},\theta_{t}),\theta_{t+1} \right\}$. 
The new Q-value combines the previous estimate with the current profit $\pi_{it}$ and the discounted continuation value, weighted by the learning rate $\alpha \in [0,1]$. The relative importance of short-term gains versus long-term continuation values is governed by the discount factor $\delta$.

The timing of events within each period is as follows.
At the beginning of period $t$, agents observe $\theta_t$, which determines the state $s_t$. They then simultaneously choose prices $p_{it}$, and payoffs $\pi_{it}$ are realized.\footnote{The action selection rule will be introduced next.}
At the beginning of period $t+1$, after observing the new demand state $\theta_{t+1}$, the next state $s'$ is determined and the Q-value corresponding to $(s_t,p_{it})$ is updated before agents select their next prices $p_{it+1}$. For all other cells in the Q-matrix, that is, for $(s,p) \neq (s_t, p_{it})$, the Q-value remains unchanged: $Q_{t+1}(s,p) = Q_t(s,p)$.

It is worth noting that Q-values are updated only after the next-period demand shock is realized. This feature implies that agents need not know the distribution of demand shocks, thereby preserving the model-free nature.

\paragraph{Action Selection}
The classic $\varepsilon$-greedy rule, widely used in reinforcement learning and commonly adopted in studies of algorithmic pricing \citep{calvano2020, klein2021, jrw2023}, is employed to determine the price chosen by each agent in each period:

\begin{equation}
p_{it}\begin{cases}
=\underset{p \in A}{\mathrm{argmax}} \ Q_{it}(s_t,p)  &\text { with probability } 1 - \varepsilon_t   \\ 
\sim \text{Uniform}(A)  &\text { with probability } \varepsilon_t
\end{cases} \label{actionmode}
\end{equation}
where $\varepsilon_t = e^{-\beta t}$ is a time-declining exploration rate that governs the trade-off between exploration and exploitation.
In each period, the agent either selects the price that maximizes its current Q-value (exploitation mode with probability $1 - \varepsilon_t$) or draws a price uniformly at random (exploration mode with probability $\varepsilon_t$).\footnote{Early in the learning process, the $\varepsilon$-greedy rule favors exploration to gather feedback from interactions with the environment. As learning progresses, it gradually shifts toward exploitation, selecting prices with the highest current Q-values.}

\paragraph{Initialization} 
The Q-matrix $\mathbf{Q}_0$ can be initialized in several ways.
In the baseline setting, $\mathbf{Q}_0$ is initialized under the assumption that the opponent sets prices completely at random.\footnote{For details of the initialization procedure, see Appendix \ref{app:initial}.}
This assumption is reasonable because, at the beginning of each session ($t = 0$), each agent operates in full exploration mode ($\varepsilon_0 = 1$) and therefore samples uniformly from the price space.
An alternative initialization is also examined in robustness checks.

\subsection{Hyperparameters}
In the baseline setting, the marginal cost for each agent is $c_i = 0$ and the i.i.d. demand shock $\theta$ takes values in $\{6, 10\}$ with equal probability.
These two demand states are denoted as low ($L$) and high ($H$), corresponding to negative and positive demand shocks, respectively.
Under these conditions, the one-shot Bertrand competition game yields equilibrium prices $p^C_L = p^C_H = 0$, whereas the monopoly prices are $p^M_L = 3$ and $p^M_H = 5$ for the low and high demand states, respectively.

The action space $A$ is discretized into $m=11$ equally spaced points over the interval $[0,5]$, such that $A = \{0, 0.5, \dots, 5\}$. Under this discretization, an additional one-shot symmetric equilibrium exists in which both agents charge a price of $0.5$ in both demand states, also corresponding to the competitive level.
For computational simplicity, the algorithm employs a one-period memory ($K = 1$).
The cardinality of the state space is therefore $\left| S \right| = 484$.
The state in period $t$ is given by $s_t = (\theta_{t-1},p_{1t-1}, p_{2t-1}, \theta_t)$.

To ensure consistent learning and sufficient exploration, I follow \citet{calvano2020} and adopt a learning rate of $\alpha = 0.15$ and an exploration rate of $\beta = 4\times10^{-6}$ in the baseline.
A broad range of $\alpha$ and $\beta$ values is also examined in robustness checks.
The discount factor $\delta$ ranges from $0.60$ to $0.99$ in increments of $0.01$, allowing for a comprehensive examination of how time preferences affect learning outcomes.
For each hyperparameter configuration $(\alpha, \beta, \delta)$, I conduct $1,000$ simulation sessions, each treated as an independent observation.

An outline of a single simulation session is given in Algorithm~\ref{alg1}.
\begin{algorithm}[h!]
\caption{Pricing Simulation Procedure for One Session}\label{alg1}
\begin{algorithmic}[1]
% \Require Price dynamics is in $G_c$
\Statex \textit{\textbf{Step 1: Initialization}}

\State $\mathbf{Q}_{i0}$ is generated

% \Comment{Discounted profit on the non-deviation path}
% \State Let one agent undercut 
% \State $(\Pi^D_1, \Pi^D_2) \gets (\pi^D_{1,1}, \pi^D_{2,1})$ 
% \Comment{Discounted profit on the deviation path}

\vspace{3mm}

\Statex \textit{\textbf{Step 2: Iteration}}
\While{convergence criterion is not satisfied}
    \State $\theta_t$ is realized
    \State $s_t=(\theta_{t-1},p_{1t-1}, p_{2t-1},\theta_t)$
    \State $p_{it}$ is determined through the action selection rule \eqref{actionmode}
    \State $\pi_i(p_{it},p_{-it},\theta_t)$ is realized
    \State $Q_{i}(s, p)$ is updated through the learning equation 
           \eqref{eq:learningfunc} \Comment{requires $\theta_{t+1}$ but is shown here for clarity}
    
    % \State $(\Pi^*_1, \Pi^*_2) \gets (\Pi^*_1+\delta^{t-1}\pi^*_{1,t}, \Pi^*_2+\delta^{t-1}\pi^*_{2,t})$ 
    % \State $(\Pi^D_1, \Pi^D_2) \gets (\Pi^D_1+\delta^{t-1}\pi^D_{1,t}, \Pi^D_2+\delta^{t-1}\pi^D_{2,t})$ 
    % \State $t \gets t+1$
\EndWhile
\end{algorithmic}
\end{algorithm}

\paragraph{Fixed-Demand Benchmark} 
I also simulate environments in which the demand state is fixed at either $L$ or $H$.\footnote{In the fixed-demand simulations, the demand state is constant, so the state variable reduces to $s_t = (p_{1t-1}, p_{2t-1})$.} Learning in these fixed-demand settings provides a natural benchmark for comparison with learning under observed demand shocks. This benchmark reflects practical situations in which firms train pricing algorithms in static environments before deploying them in dynamic markets subject to stochastic demand shocks.

\paragraph{Convergence} 
Although convergence of Q-learning has been proven in single-agent settings under specific conditions \citep{wd1992Q}, it is not theoretically guaranteed in strategically interdependent environments.
Following \citet{calvano2020}, I employ an empirical convergence criterion: learning is considered complete when each agent's optimal strategy remains unchanged for $100,000$ consecutive periods.\footnote{Specifically, convergence is achieved when, for each agent $i$ and state $s$, the optimal price $p_{it}(s) = \underset{p \in A}{\mathrm{argmax}}\,\, Q_{i}(s, p)$ remains constant over $100,000$ consecutive iterations.}
The simulation terminates once this criterion is satisfied or after one billion iterations, whichever occurs first.
Under the baseline hyperparameters $\alpha = 0.15$ and $\beta = 4 \times 10^{-6}$, the average number of iterations required for convergence is $2,677,436$.\footnote{Convergence under fixed demand states $L$ and $H$ requires $1,642,801$ and $1,815,529$ iterations, respectively. These smaller numbers are expected, as the state space in each single-demand simulation is only half the size of the state space under observed demand shocks.}

\subsection{Theoretical Predictions} 
\begin{figure}[H]
    \centering
    \includegraphics[width=0.99\textwidth]{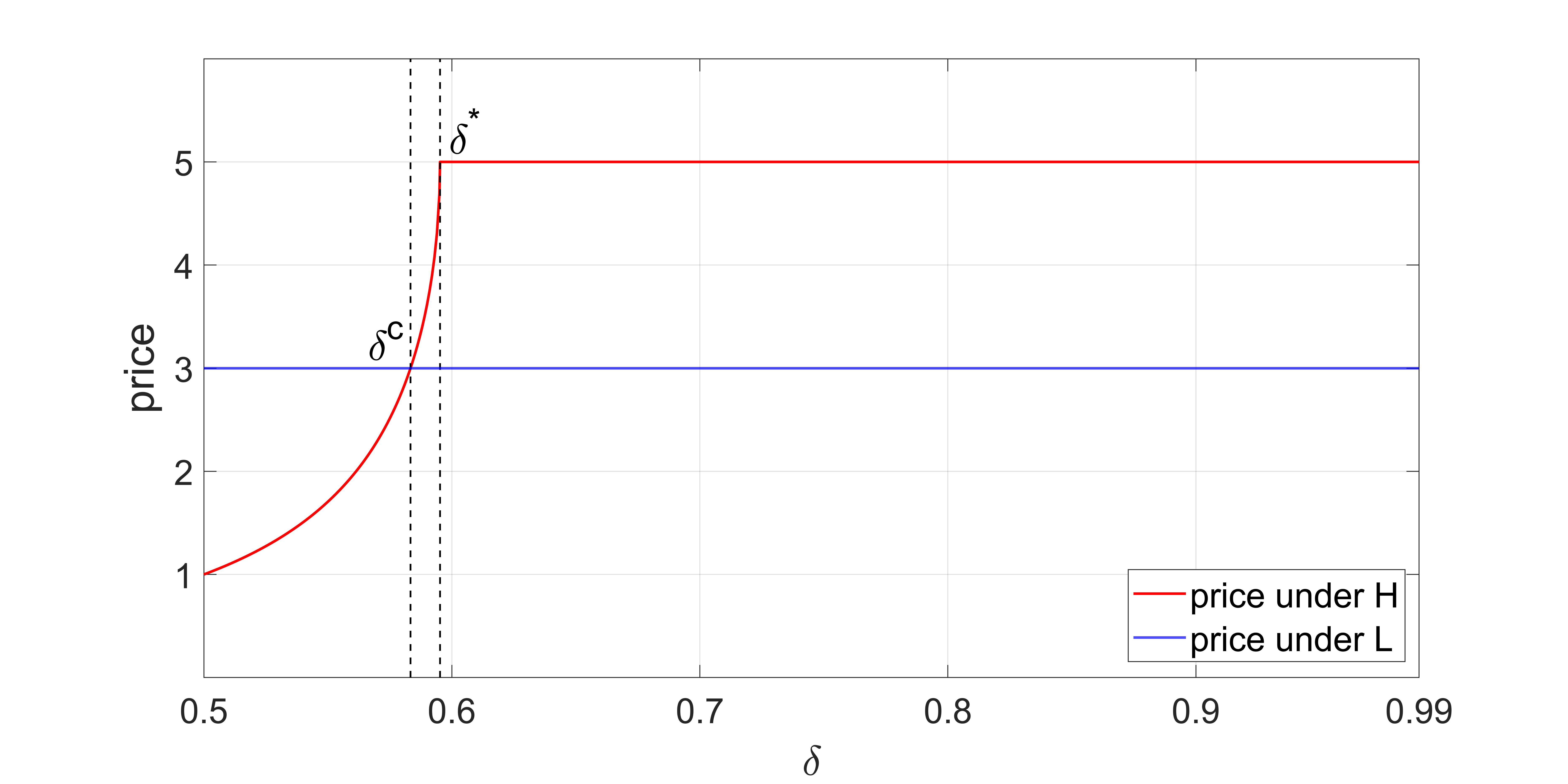}
    \caption{Predicted Prices for Both Demand States across $\delta$}
    \label{fig:theoretical}
\end{figure}

I outline the theoretical predictions for pricing patterns in the most collusive equilibrium that can be sustained at each value of $\delta$, assuming firms employ the grim-trigger strategy.\footnote{For a formal theoretical analysis, see \citet{rs1986}.}
Figure~\ref{fig:theoretical} illustrates the predicted pricing patterns as a function of $\delta$.
Under $L$, the monopoly price of $3$ is sustainable whenever $\delta \ge 0.5$. 
Under $H$, charging the monopoly price of $5$ is feasible only when $\delta \ge \delta^* = \tfrac{25}{42} \approx 0.595$.
As $\delta$ decreases further, a second cutoff, $\delta^c = \tfrac{7}{12} \approx 0.583$, marks the point where the cyclicality of pricing reverses.
When $\delta > \delta^c$, the pricing pattern is procyclical, with higher prices sustained under $H$.
When $\delta < \delta^c$, pricing becomes countercyclical: firms sustain lower prices under $H$, despite higher profits in that demand state.

The intuition behind the cyclicality of pricing is as follows. Deviation incentives are always stronger under $H$ because the short-term gain from undercutting increases with market size. 
Thus, the best collusive price under $L$, namely charging the monopoly price of $3$, can always be sustained as long as $\delta \ge 0.5$.
However, the highest sustainable collusive price under $H$ depends on the trade-off between the short-term deviation incentive and the expected long-term continuation value, which is governed by $\delta$.
When $\delta$ is sufficiently high, the incentive constraint holds under $H$, allowing firms to sustain higher prices than $3$ and generate procyclical pricing. As $\delta$ falls, firms must lower prices under $H$ to mitigate deviation incentives, and this adjustment eventually gives rise to countercyclical pricing.

\begin{flushleft}
\textbf{Prediction 1:} 
When $\delta > \delta^c$, the pricing pattern in the most collusive equilibrium is procyclical.
\end{flushleft}

\begin{flushleft}
\textbf{Prediction 2:} 
When $0.5 \leq \delta < \delta^c$, the pricing pattern in the most collusive equilibrium is countercyclical.
\end{flushleft}

\begin{flushleft}
\textbf{Prediction 3:} 
When $\delta < 0.5$, the pricing pattern is rigid, with both firms charging prices at the competitive level across demand states.
\end{flushleft}

%%%%%%%%%%%%%%%%%%%%%%%%%%%%%%%%%%%%%%%%%%%%%%%%%%%%%%%

\section{Pricing Patterns}\label{sec:pattern}
\subsection{Derivation of Pricing Patterns}
To identify pricing patterns, I first obtain the limit strategies.
For each agent $i$, the limit strategy is the optimal policy derived from the converged Q-matrix, which maps every state to its optimal price, formally defined as
\begin{equation}
    p_i^{*}(s) = \underset{p \in A}{\mathrm{argmax}}\,\, Q_{i}(s, p), \quad \forall s \in S \label{optimal_policy}
\end{equation}

Given the limit strategies of both agents, the induced price dynamics can be represented by a directed graph, where each node corresponds to a demand state and a price pair, and each directed link denotes a possible transition from one node to another, as implied by the limit strategies and the stochastic evolution of demand.
In the long run, the price dynamics converge to an absorbing price cycle, i.e., once they enter this cycle, all subsequent transitions remain within it. This price cycle defines a finite Markov process, in which each node is visited with a stationary (steady-state) probability. Using the stationary distribution, I compute the average long-run price conditional on each demand state. The formal construction and derivation are provided in the Appendix \ref{app:steady}. Based on these steady-state averages, the resulting pricing patterns are defined as follows.

\begin{flushleft}
\textbf{Definition 1:} 
A price cycle exhibits symmetric and rigid pricing (Sym-Rigid) if it consists of exactly two nodes and maintains identical prices across both agents and demand states.
\end{flushleft}

\begin{flushleft}
\textbf{Definition 2:} 
A price cycle exhibits procyclical pricing (Pro-Cycle) if the average long-run prices for both agents are strictly higher under $H$ than under $L$.
\end{flushleft}

\begin{flushleft}
\textbf{Definition 3:} 
A price cycle exhibits countercyclical pricing (Counter-Cycle) if the average long-run prices for both agents are strictly higher under $L$ than under $H$.
\end{flushleft}

Pricing patterns that do not satisfy any of the above definitions are classified as \textit{Others}.
The two-node constraint is imposed on Sym-Rigid because allowing more than two nodes would necessarily generate price variation across demand states, thereby violating the notion of price rigidity.
In contrast, the characterization of Pro-Cycle and Counter-Cycle is based on average long-run prices and does not depend on the number of nodes.
This definition allows for local price variation while still capturing the fundamental relationship between prices and demand conditions.
Both Pro-Cycle and Counter-Cycle are therefore referred to as forms of demand-contingent pricing.

Table \ref{table:pricing_pattern} summarizes the definition of each pricing pattern.

\begin{table}[h!]
\centering
\scalebox{0.99}{
\begin{threeparttable}
\caption{Definitions of Pricing Patterns}
\label{table:pricing_pattern}
\def\arraystretch{1.5}
\begin{tabular}{ccc}
\hline
Name & Abbreviation & Price Relations \\ \hline
Symmetric and Rigid Pricing & Sym-Rigid & 
$p_L^1=p_L^2=p_H^1=p_H^2$ (two nodes) \\[2pt]
Procyclical Pricing & Pro-Cycle & 
$p_H^1>p_L^1 \text{ and } p_H^2>p_L^2$ \\[2pt]
Countercyclical Pricing & Counter-Cycle & 
$p_L^1>p_H^1 \text{ and } p_L^2>p_H^2$ \\ \hline
\end{tabular}
\begin{tablenotes}
  \small
  \item Notes: $p^i_\theta$ denotes agent $i$’s average long-run price in demand state $\theta$, computed from the stationary distribution of the price cycle.
\end{tablenotes}
\end{threeparttable}}
\end{table}

\begin{figure}[h!]
\centering
\begin{subfigure}{0.40\textwidth}
    \includegraphics[width=\textwidth]{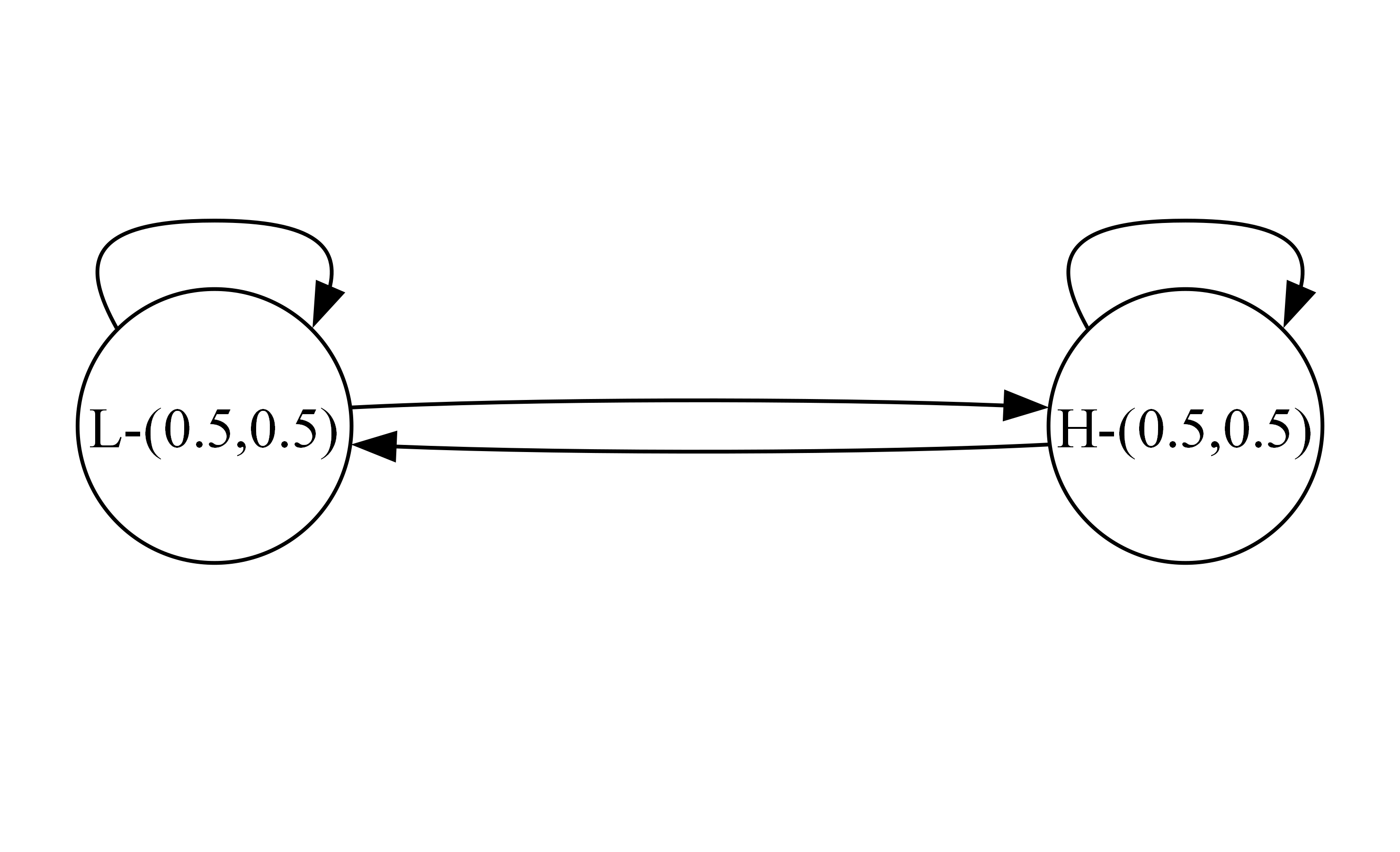}
    \centering
    \caption{Sym-Rigid ($\delta=0.52$)}
    \label{fig:exampleG_1}
\end{subfigure}
\hfill
\begin{subfigure}{0.48\textwidth}
    \includegraphics[width=\textwidth]{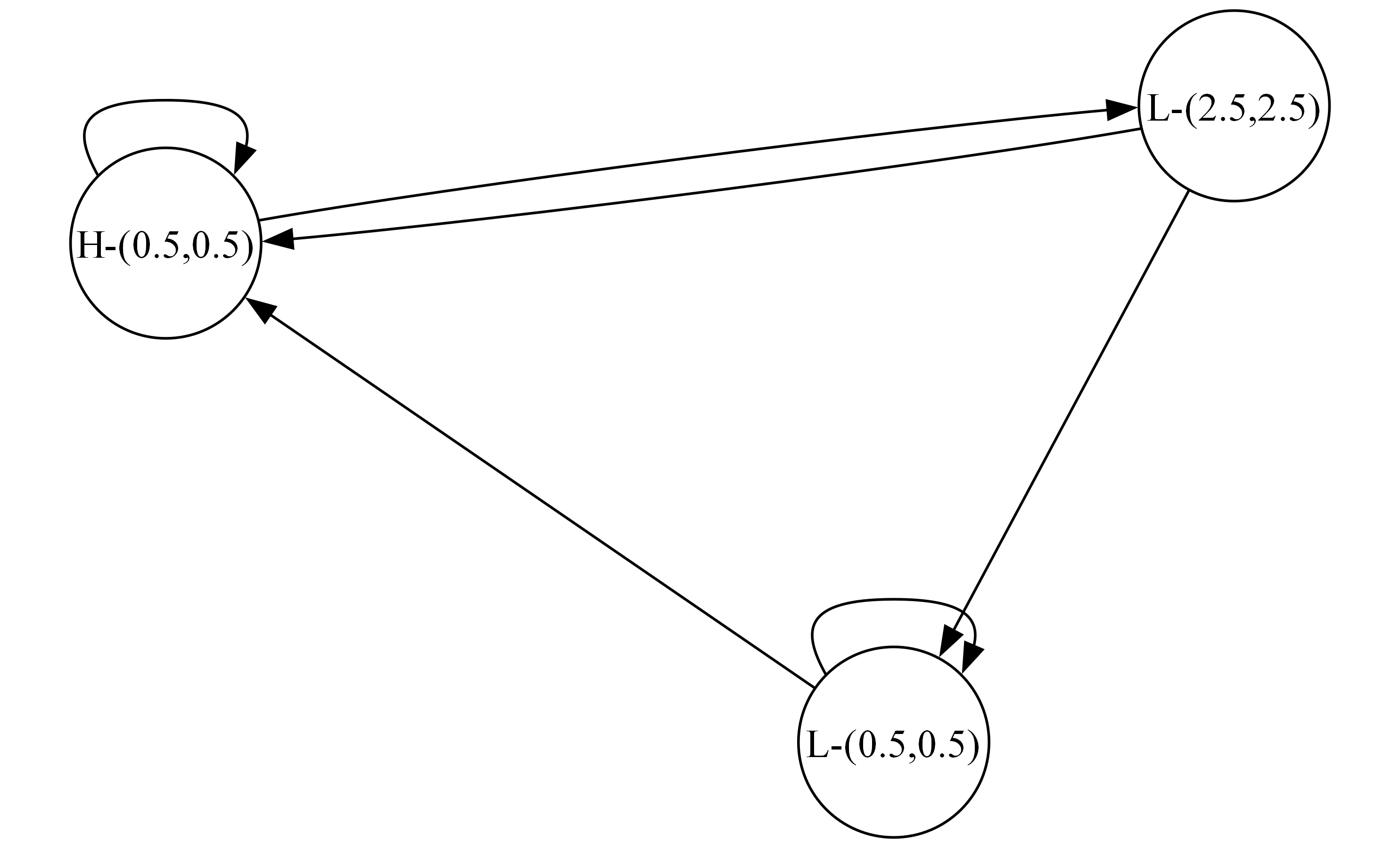}
    \centering
    \caption{Counter-Cycle ($\delta=0.62$)}
    \label{fig:exampleG_2}
\end{subfigure}
\hspace{2cm}
\begin{subfigure}{0.48\textwidth}
    \includegraphics[width=\textwidth]{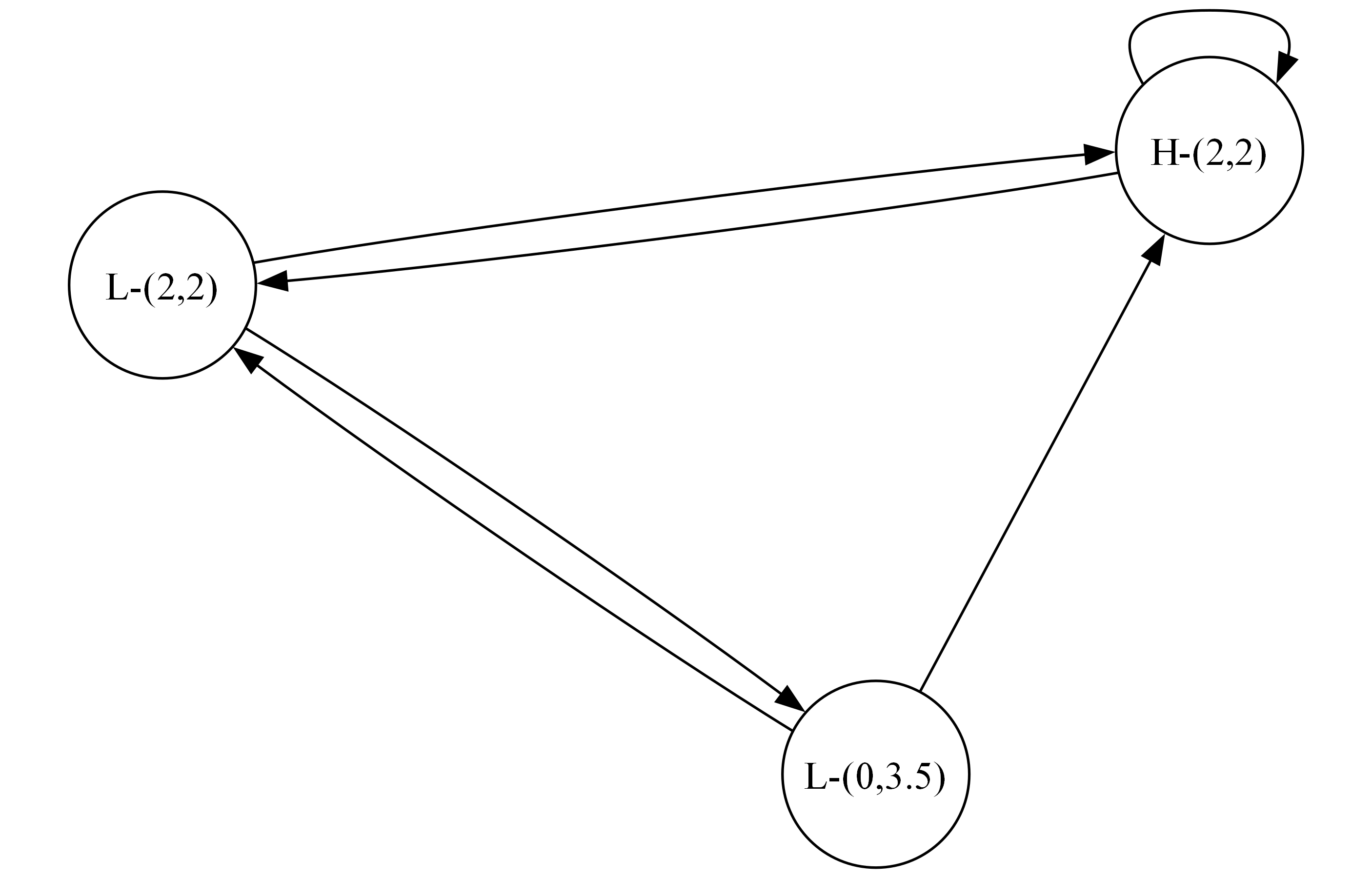}
    \centering
    \caption{Others ($\delta=0.84$)}
    \label{fig:exampleG_3}
\end{subfigure}
\begin{subfigure}{0.48\textwidth}
    \includegraphics[width=\textwidth]{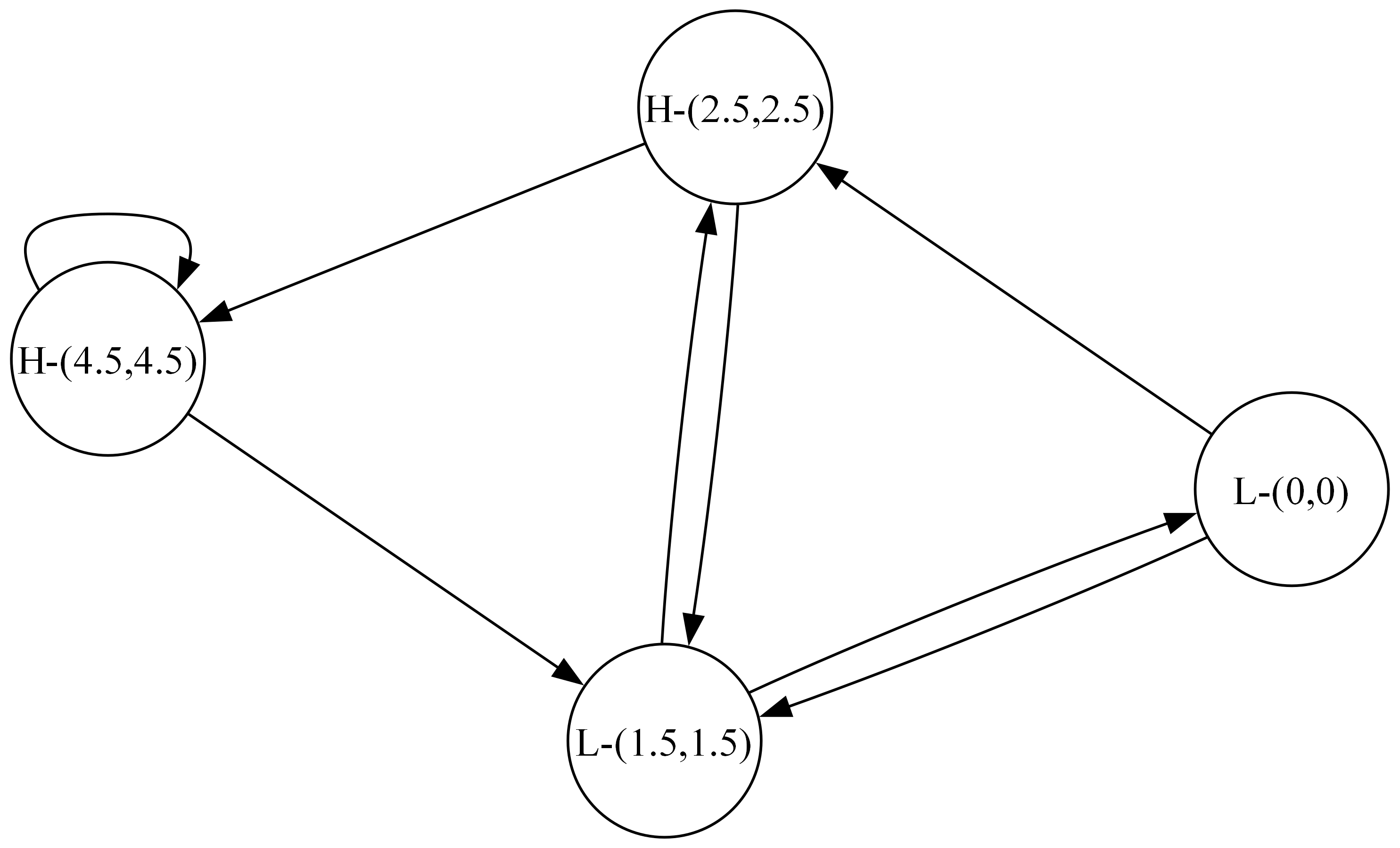}
    \centering
    \caption{Pro-Cycle ($\delta=0.96$)}
    \label{fig:exampleG_4}
\end{subfigure}
\caption{Examples of Price Cycles across Different Pricing Patterns}
\label{fig:examplesG}
\end{figure}

Figure \ref{fig:examplesG} illustrates examples of price cycles across different categories.
The label within each node indicates the demand state and the corresponding price pair.
Directed edges represent stochastic transitions between nodes, while an arrow looping back to the same node represents a self-loop conditional on remaining in the same demand state.
For example, in Figure \ref{fig:exampleG_1}, each node has two outgoing edges: one leading to the other node and one self-loop.
This structure shows that both agents charge a price of $0.5$ regardless of demand shocks, forming Sym-Rigid.

\subsection{Distribution of Pricing Patterns}
\begin{figure}[h!]
    \centering
    \includegraphics[width=0.7\textwidth]{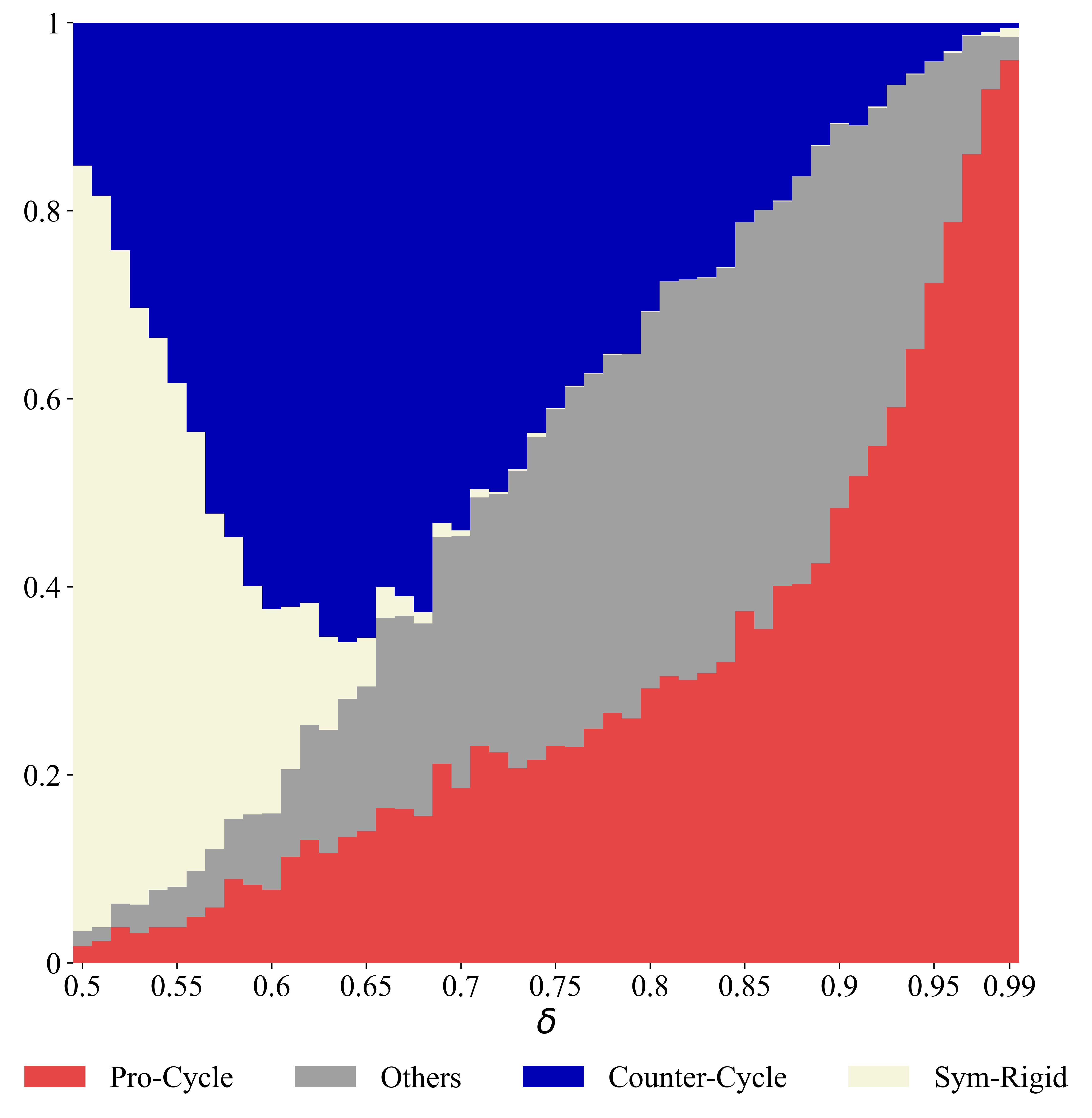}
    \caption{Distribution of Pricing Patterns across $\delta$}
    \label{fig:distribution_pattern}
\end{figure}

Figure \ref{fig:distribution_pattern} presents the distribution of pricing patterns as a function of $\delta$.\footnote{For detailed price dynamics across $\delta$, see Figure \ref{fig:price_pattern_delta} in the Appendix.}
When $\delta$ is close to $0.5$, Sym-Rigid dominates, with agents uniformly charging a price of $0.5$ at the competitive level. 
As $\delta$ increases, the share of Counter-Cycle rises sharply, peaking above $50\%$ between $\delta = 0.57$ and $\delta = 0.70$, and then declining thereafter.
In contrast, the prevalence of Pro-Cycle increases monotonically and accelerates at high $\delta$, surpassing $50\%$ when $\delta > 0.90$ and approaching $100\%$ at $\delta = 0.99$.
In the intermediate range of $\delta$, a non-negligible fraction of sessions falls into the category of Others.
Others typically combines features of Pro-Cycle and Counter-Cycle, with two agents charging higher prices in different demand states.\footnote{Figure \ref{fig:exampleG_3} provides an example.}
This pattern indicates that, at intermediate levels of $\delta$, the balance between short-term and long-term incentives does not fully determine the learned pricing pattern, leaving some price cycles without a consistent demand-contingent structure because two agents respond differently to demand shocks.
% results
Overall, the observed dominance regions of different pricing patterns across $\delta$ are qualitatively consistent with \citet{rs1986}.\footnote{A distinctive feature of Q-learning agents, however, is their gradual and adaptive learning process, as reflected in the coexistence of multiple pricing patterns at the same $\delta$. This contrasts with the theoretical prediction that agents should converge directly to a unique equilibrium pricing pattern for each $\delta$.}

\begin{flushleft}
\textbf{Result 1:} 
Under observed demand shocks, Q-learning agents learn distinct pricing patterns across $\delta$: procyclical pricing dominates at high $\delta$, countercyclical pricing dominates at low $\delta$, and uniformly low prices at the competitive level prevail when $\delta$ is close to $0.5$.
\end{flushleft}

%%%%%%%%%%%%%%%%%%%%%%%%%%%%%%%%%%%%%%%%%%%%%%%%%%%%%%%
\subsection{Evaluation of Pricing Patterns}  \label{sec:procycle}
After outlining the overall distribution of pricing patterns, 
I now evaluate the two demand-contingent pricing patterns, Pro-Cycle and Counter-Cycle.

\subsubsection{Procyclical Pricing}  \label{sec:procycle}

\begin{figure}[h!]
\centering
\begin{subfigure}{0.48\textwidth}
    \includegraphics[width=\textwidth]{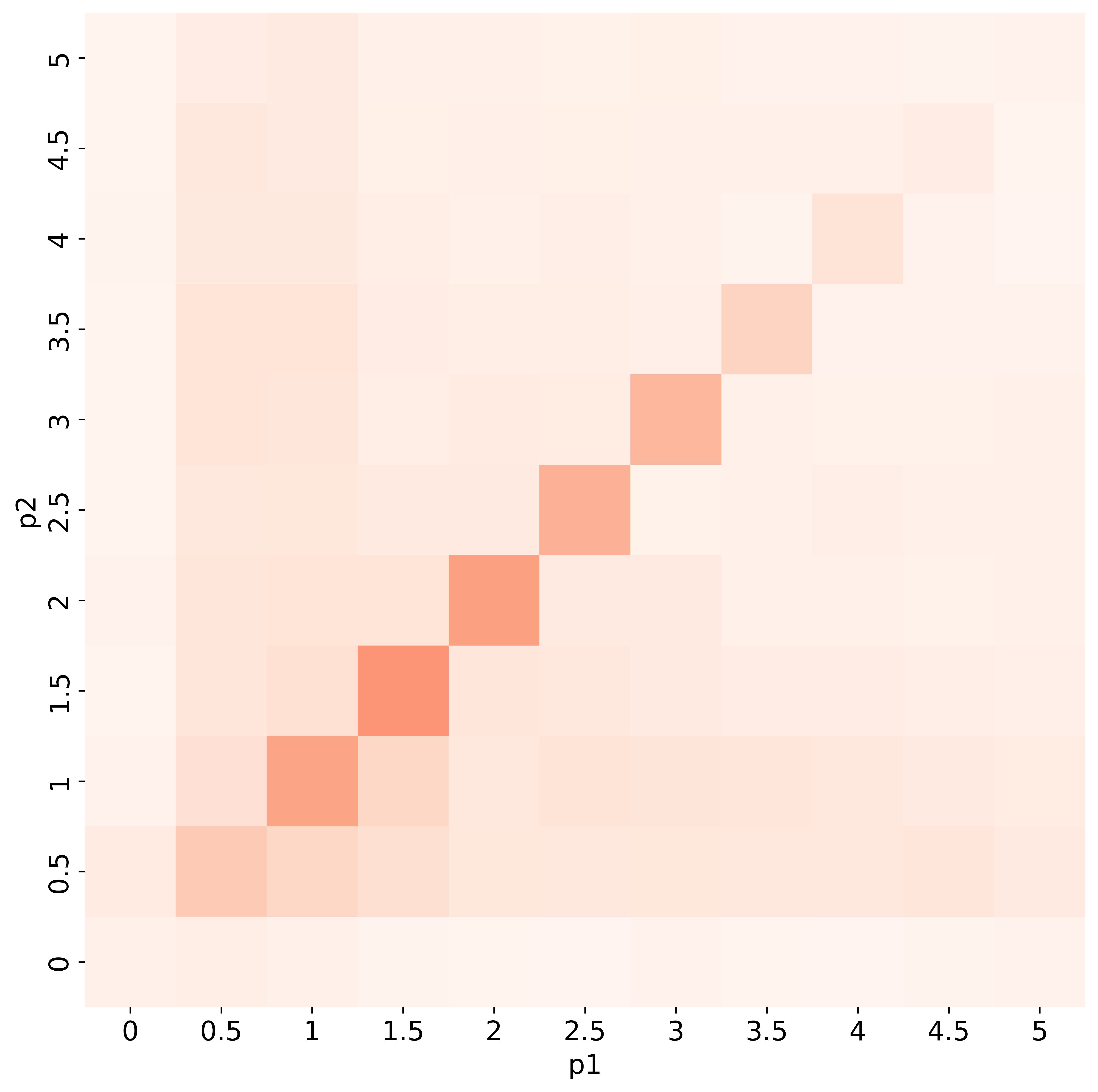}
    \caption{Price Pairs under L}
    \label{fig:pricedist_pc1}
\end{subfigure}
\hfill
\begin{subfigure}{0.48\textwidth}
    \includegraphics[width=\textwidth]{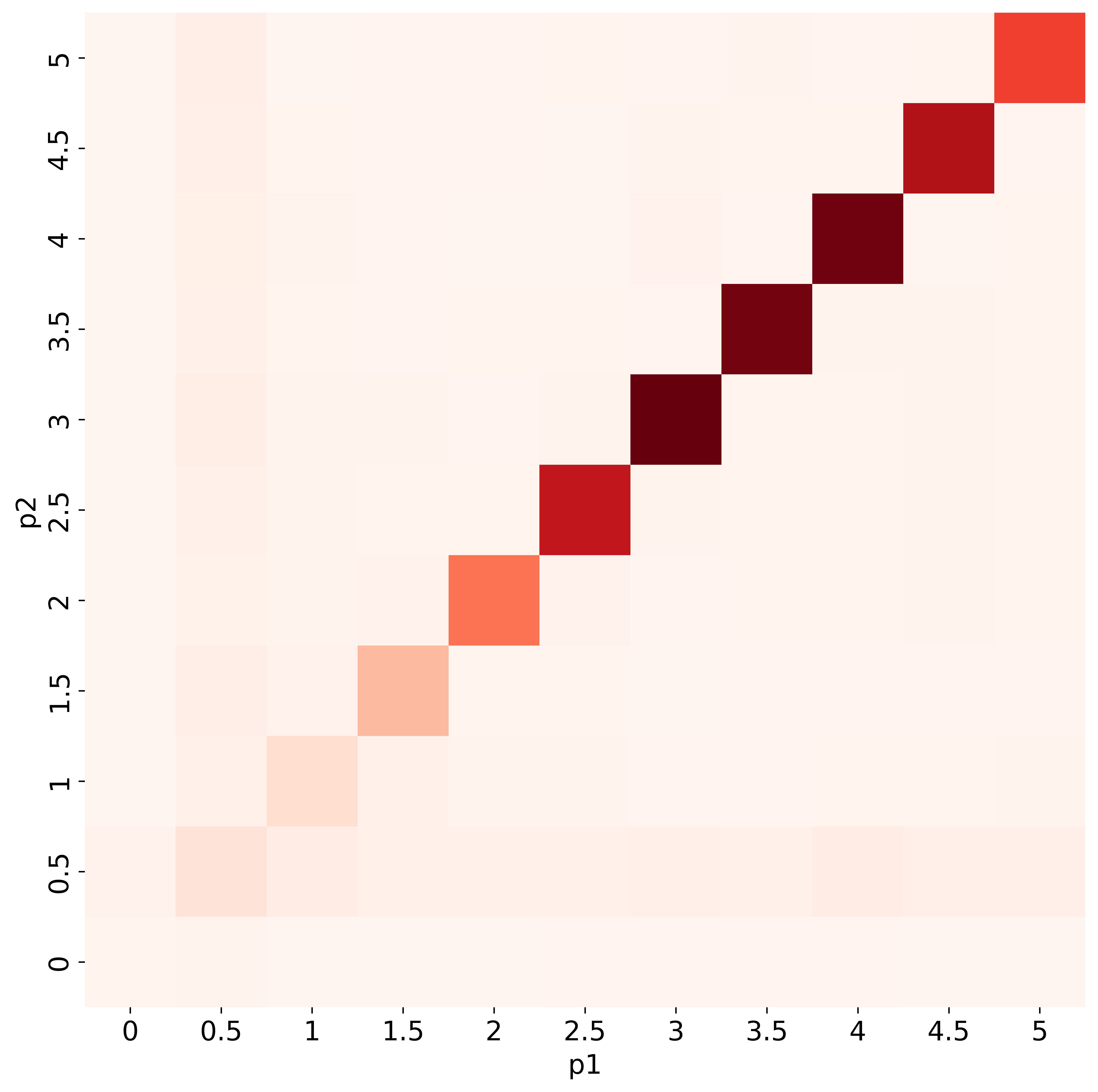}
    \caption{Price Pairs under H}
    \label{fig:pricedist_pc2}
\end{subfigure}
\caption{Stationary Probability of Price Pairs under Pro-Cycle ($\delta = 0.96$)}
\caption*{Notes: Each panel presents the cross-session average stationary probability of each price pair, separately for $L$ and $H$. Darker colors indicate higher stationary probabilities.}
\label{fig:pricedist_pc}
\end{figure}

\paragraph{Graph Structure} 
To investigate how price nodes are distributed within the price cycle, I compute the stationary probability of each node, which directly measures how frequently it is visited in the long run. 
Figure \ref{fig:pricedist_pc} presents two heatmaps displaying the cross-session average stationary probability of each price pair, shown separately for $L$ and $H$, under Pro-Cycle at $\delta = 0.96$.\footnote{
I first compute the stationary probability of each price node within each session and then average these probabilities across sessions.
} Darker colors indicate higher probabilities.

The heatmaps reveal two salient pricing features. First, although occasional asymmetric pricing occurs, symmetric pricing predominates, with symmetric price pairs consistently being supracompetitive in both demand states. This predominant symmetry suggests that Q-learning agents have largely learned to coordinate their pricing, though coordination remains imperfect.
Second, in both demand states, prices exhibit a distribution rather than a single point, which reflects stochasticity in the learning process. Sustained prices are concentrated within a narrow range of higher prices under $H$, whereas prices under $L$ display a wider dispersion.\footnote{Supracompetitive pricing sustained through a stationary distribution that assigns high probability to high-price outcomes represents another form of persistence \citep{abada2025algorithmic}.}

\paragraph{Performance} 
Table \ref{table:prices96} reports each agent's prices and profits at $\delta = 0.96$ across different demand environments.
Panel A shows that under observed demand shocks, the dominant pricing pattern of Pro-Cycle accounts for about $80\%$ of the sessions. 
Average prices are $2.14$ under $L$ and $3.13$ under $H$, representing $71\%$ and $63\%$ of the respective monopoly prices. 
The expected profit across demand states exceeds $70\%$ of the best collusive profit (i.e., the equally split monopoly profit), thus yielding supracompetitive outcomes.\footnote{Figure \ref{fig:profit_dynamics} further shows that the profitability of Pro-Cycle increases with $\delta$.}

\begin{table}[h!]
\centering
\scalebox{0.99}{
\begin{threeparttable}
\caption{Summary Statistics for Pricing Patterns at $\delta=0.96$} 
\label{table:prices96}
\begin{tabular}{ccccccc}
\hline
Category  & Ratio\textsuperscript{a}      & $\mathrm{p_L}$ & $\mathrm{\pi_L}$ & $\mathrm{p_H}$ & $\mathrm{\pi_H}$ & Expected Profit\textsuperscript{b} \\ \hline
\multicolumn{7}{l}{\textit{Panel A: Observed  Demand Shocks}}                                                                              \\
Pro-Cycle\textsuperscript{c} & 0.79  & 2.14  & 2.76   & 3.13  & 9.47   & 6.12                                                                      \\
          &       & (0.71\textsuperscript{d})  & (0.61)   & (0.63)  & (0.76)   & (0.72)                                                                      \\
          &       & [1.47\textsuperscript{e}]  &        & [2.93]  &        &                                                                           \\
          &       &       &        &       &        &                                                                           \\
\multicolumn{7}{l}{\textit{Panel B: Fixed-Demand Benchmark}}                                                                           \\
Sym-1Node (L)\textsuperscript{f} & 0.94  & 2.14  & 4.04   &       &        & \multirow{4}{*}{\begin{tabular}[c]{@{}c@{}} 7.26\\      (0.85)\end{tabular}} \\
          &       & (0.71)  & (0.90)    &       &        &                                                                           \\
Sym-1Node (H) & 0.97  &       &        & 3.09  & 10.48  &                                                                           \\
          &       &       &        & (0.62)  & (0.84)   &                                                                           \\ \hline
\end{tabular}
\begin{tablenotes}
  \small
  \item Notes: a. Frequency of the corresponding pricing pattern among all sessions. 
  \item b. Average profit across both demand states.
  \item c. Pro-Cycle may involve asymmetric pricing. However, as average prices and profits are nearly identical for both agents, Panel A shows only agent 1's results.
  \item d. Proportion relative to the monopoly price or the equally split monopoly profit.
  \item e. Average effective market price.
  \item f. Sym-1Node denotes the pricing pattern under the fixed-demand benchmark, where the price cycle contains exactly one node, and both agents charge the same price.
\end{tablenotes}
\end{threeparttable}} 
\end{table}

To further evaluate Pro-Cycle, I examine the pricing behavior of Q-learning agents under the fixed-demand benchmark (fixing either $L$ or $H$).
Panel B reports the pricing pattern, denoted as \textit{Sym-1Node}, for each fixed demand state. Sym-1Node refers to a price cycle consisting of exactly one node, where both agents charge the same price.\footnote{Furthermore, under Sym-1Node, each node includes only the price pair, without the demand state, since demand is fixed.} Under Sym-1Node, agents on average charge prices of $2.14$ in $L$ and $3.09$ in $H$, which are close to those observed under Pro-Cycle.
Nevertheless, the average profit across the two fixed-demand environments is $85\%$, compared with $72\%$ under Pro-Cycle. 
The lower profits under Pro-Cycle are consistent with the lower effective market prices observed in both demand states ($1.47$ under $L$ and $2.93$ under $H$). These lower effective prices result from asymmetric pricing (see Figure \ref{fig:pricedist_pc}), where an agent occasionally undercuts or is undercut by its rival. The gains from undercutting do not fully offset the losses from being undercut, leading to an overall decline in expected profits.

\begin{flushleft}
\textbf{Result 2:} When $\delta$ is high, the predominant pricing pattern is Pro-Cycle. It exhibits a distribution over supracompetitive prices, with a tighter concentration under $H$. This pricing pattern yields supracompetitive profits, and both its prevalence and profitability increase with $\delta$.
\end{flushleft}

%%%%%%%%%%%%%%%%%%%%%%%%%%%%%%%%%%%%%%%%%%%%%%%%%
\subsubsection{Countercyclical Pricing}  \label{sec:Countercyclical}
\begin{figure}[h!]
\centering
\begin{subfigure}{0.48\textwidth}
    \includegraphics[width=\textwidth]{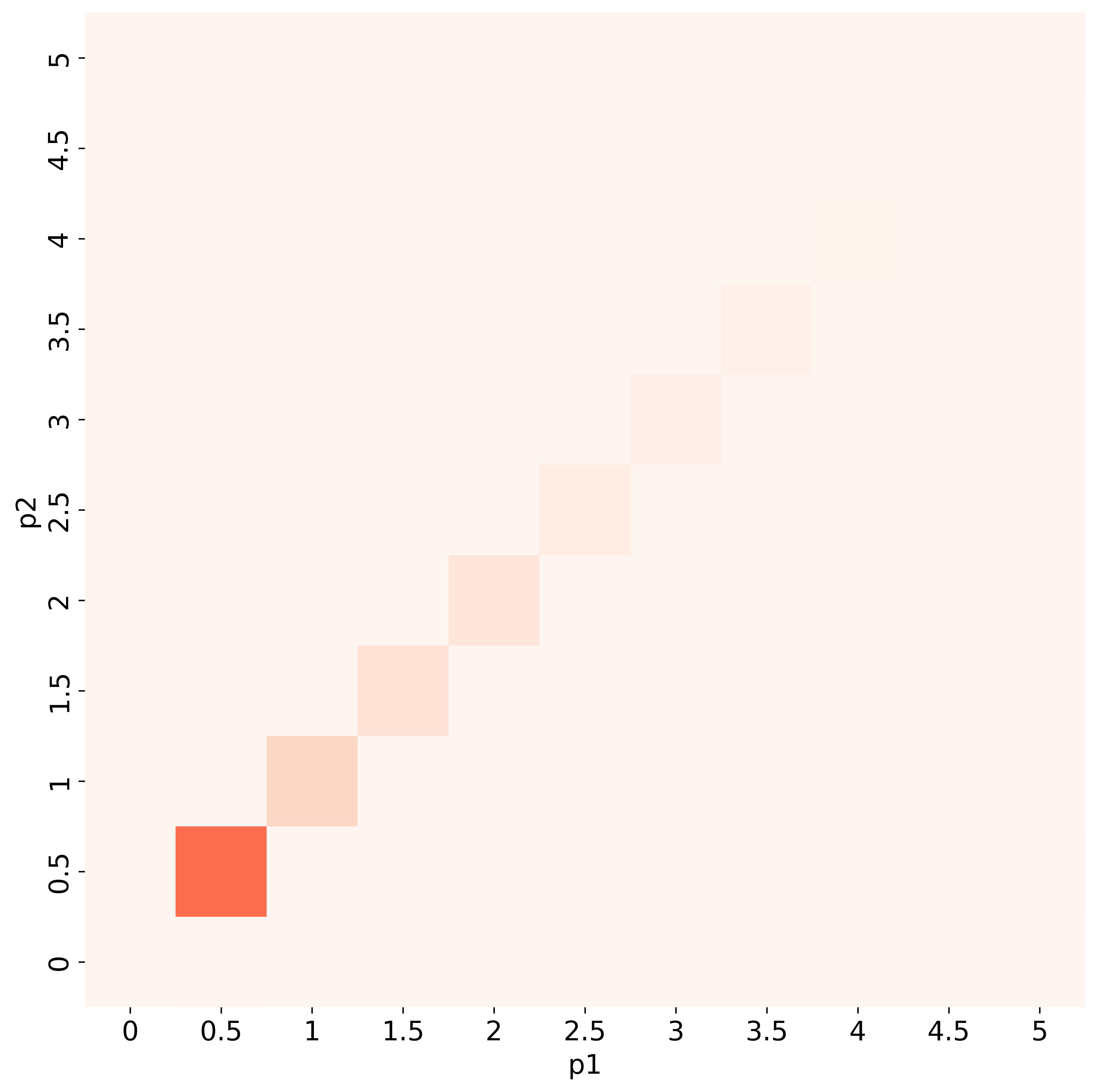}
    \caption{Price Pairs under L}
    \label{fig:pricedist_cc1}
\end{subfigure}
\hfill
\begin{subfigure}{0.48\textwidth}
    \includegraphics[width=\textwidth]{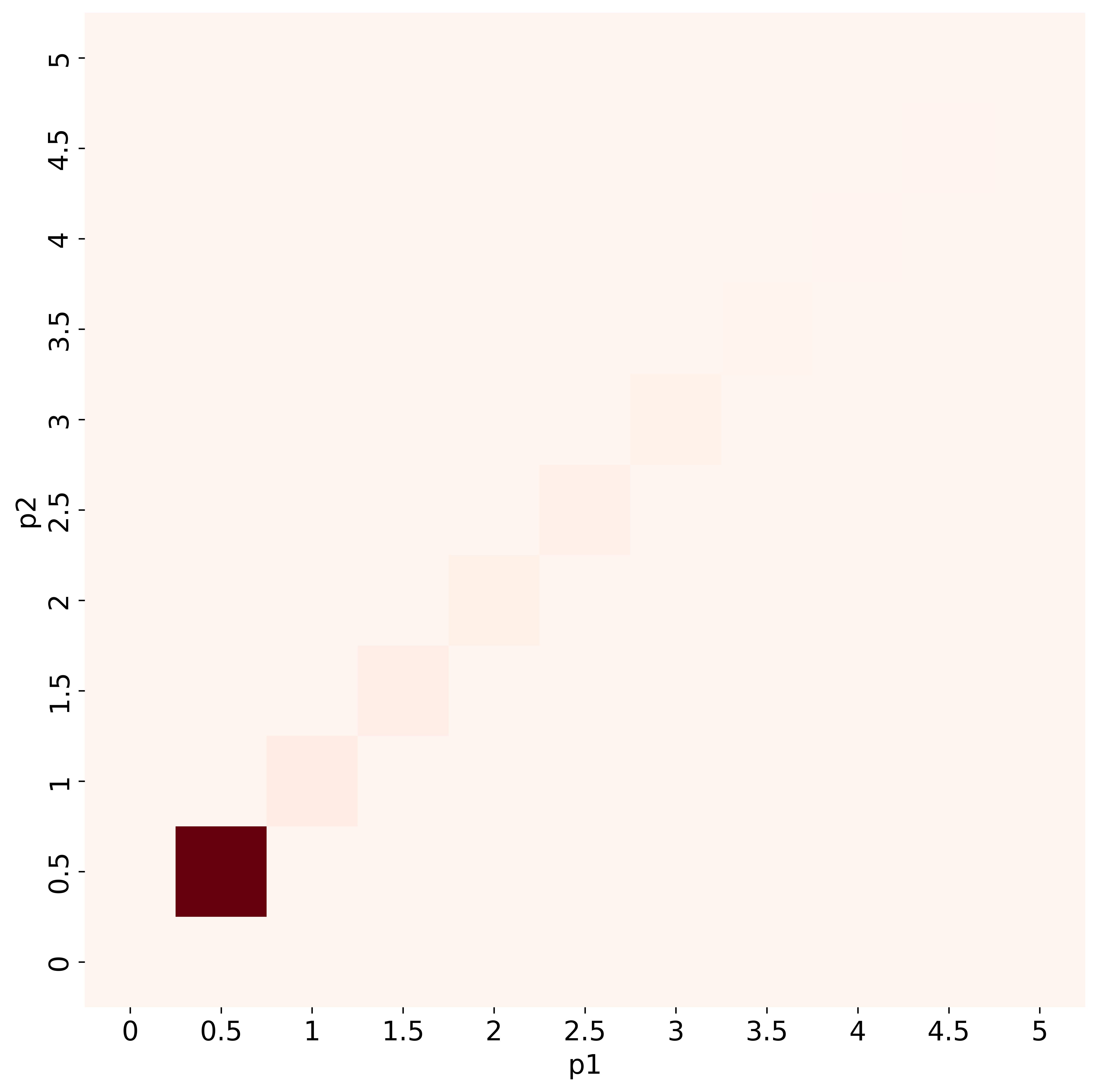}
    \caption{Price Pairs under H}
    \label{fig:pricedist_cc2}
\end{subfigure}
\caption{Stationary Probability of Price Pairs under Counter-Cycle ($\delta = 0.66$)}
\caption*{Notes: Each panel presents the cross-session average stationary probability of each price pair, separately for $L$ and $H$. Darker colors indicate higher stationary probabilities.}
\label{fig:pricedist_cc}
\end{figure}

\paragraph{Graph Structure} 
Figure \ref{fig:pricedist_cc} displays the cross-session average stationary probability of each price pair under Counter-Cycle at $\delta = 0.66$. Compared with $\delta = 0.96$, the assymetric pricing observed at high $\delta$ almost completely disappears, and the stationary mass shifts toward a more concentrated subset of low-price nodes. This occurs because, as $\delta$ decreases, Q-learning agents place greater weight on immediate rewards relative to continuation values, making high-price coordination unsustainable and effectively eliminating asymmetric pricing as agents learn to avoid being undercut.

Within this concentrated distribution, the price pair $(0.5, 0.5)$ has the highest stationary mass in both demand states, and its prominence is especially pronounced under $H$, roughly twice that under $L$.
This prominence of $H-(0.5,0.5)$  arises from two sources that contribute almost equally: (i) a high self-loop tendency, and (ii) substantial inflows from other nodes.\footnote{Let $P$ denote the transition matrix and $\psi^*$ the corresponding stationary distribution. For any node $j$, its stationary probability $\psi^*_j$ can be decomposed as
$$
\psi^*_j=\psi^*_j P_{j j}+\sum_{i \neq j} \psi^*_i P_{i j}=\psi^*_j P_{j j}+\psi^*_j (1-P_{j j})
$$
where the first term represents the self-loop and the second term represents inflows from other nodes. When the node has a self-loop, $P_{jj}=0.5$, so the two components are exactly equal; otherwise the entire stationary mass comes from inflows. Because $H-(0.5,0.5)$ appears in nearly all sessions ($99.8\%$) and has a high self-loop rate ($89\%$), the two contributions on average are approximately balanced.}
Consequently, prices stabilize at the competitive level under $H$, in contrast to the greater price variation under $L$, thereby generating countercyclical pricing and echoing the price wars during booms described by \cite{rs1986}.

\paragraph{Performance} 
Table \ref{table:prices66} reports each agent's prices and profits at $\delta = 0.66$ across different demand environments.
Panel A shows that under Counter-Cycle, which accounts for $60\%$ of the sessions, the average price under $L$ is $1.44$ ($48\%$ of the monopoly price), yielding $53\%$ of the equally split monopoly profit, whereas under $H$ it falls to $0.77$ ($15\%$), corresponding to only $26\%$.
Overall, the expected profit amounts to $33\%$ of the best collusive profit, as profit gains under $L$ are largely offset by weak profitability under $H$.

\begin{table}[h!]
\centering
\scalebox{0.99}{
\begin{threeparttable}
\caption{Summary Statistics of Pricing Patterns at $\delta=0.66$} 
\label{table:prices66}
\begin{tabular}{ccccccc}
\hline
Category      & Ratio      & $\mathrm{p_L}$ & $\mathrm{\pi_L}$ & $\mathrm{p_H}$ & $\mathrm{\pi_H}$ & Expected Profit \\ \hline
\multicolumn{7}{l}{\textit{Panel A: Observed   Demand Shocks}}                                                                                  \\
Counter-Cycle & 0.60  & 1.44  & 2.39   & 0.77  & 3.25   & 2.82                                                                      \\
              &       & (0.48)  & (0.53)   & (0.15)  & (0.26)   & (0.33)                                                                      \\
              &       & [1.23]  &        & [0.76]  &        &                                                                           \\
              &       &       &        &       &        &                                                                           \\
\multicolumn{7}{l}{\textit{Panel B: Fixed-Demand Benchmark}}                                                                               \\
Sym-1Node (L)     & 0.88  & 0.51  & 1.39   &       &        & \multirow{4}{*}{\begin{tabular}[c]{@{}c@{}}2.34\\      (0.27)\end{tabular}} \\
              &       & (0.17)  & (0.31)   &       &        &                                                                           \\
Sym-1Node (H)     & 0.81  &       &        & 0.73  & 3.28   &                                                                           \\
              &       &       &        & (0.15)  & (0.26)   &                                                                           \\ \hline
\end{tabular}
\begin{tablenotes}
  \small
  \item Notes: The settings are identical to those in Table \ref{table:prices96}. As prices and profits are nearly identical for agents 1 and 2, results are reported for agent 1 only.
\end{tablenotes}
\end{threeparttable}} 
\end{table}

For comparison, Panel B shows that under fixed demand states, Q-learning agents charge prices close to the competitive level, $0.51$ in $L$ and $0.73$ in $H$. Although profits under fixed demand $H$ are comparable to those achieved during booms, markedly weak performance under fixed demand $L$ leads to lower overall profitability than under Counter-Cycle. This pattern contrasts with the case at $\delta = 0.96$, where the dominant pricing pattern, Pro-Cycle, yields lower expected profits. Hence, relative to the fixed-demand benchmark, learning under observed demand shocks reverses its effect on profitability, being higher at $\delta = 0.66$ but lower at $\delta = 0.96$.

\begin{flushleft}
\textbf{Result 3:} 
When $\delta$ is low, the predominant pricing pattern is Counter-Cycle. It exhibits sharply reduced price dispersion, with prices concentrated near the competitive level and tighter concentration under $H$. It yields supracompetitive profits under $L$ but performs poorly under $H$, leaving overall expected profits relatively low.
\end{flushleft}

\subsection{Profit Dynamics}  
\begin{figure}[h!]
\centering
\includegraphics[width=0.7\textwidth]{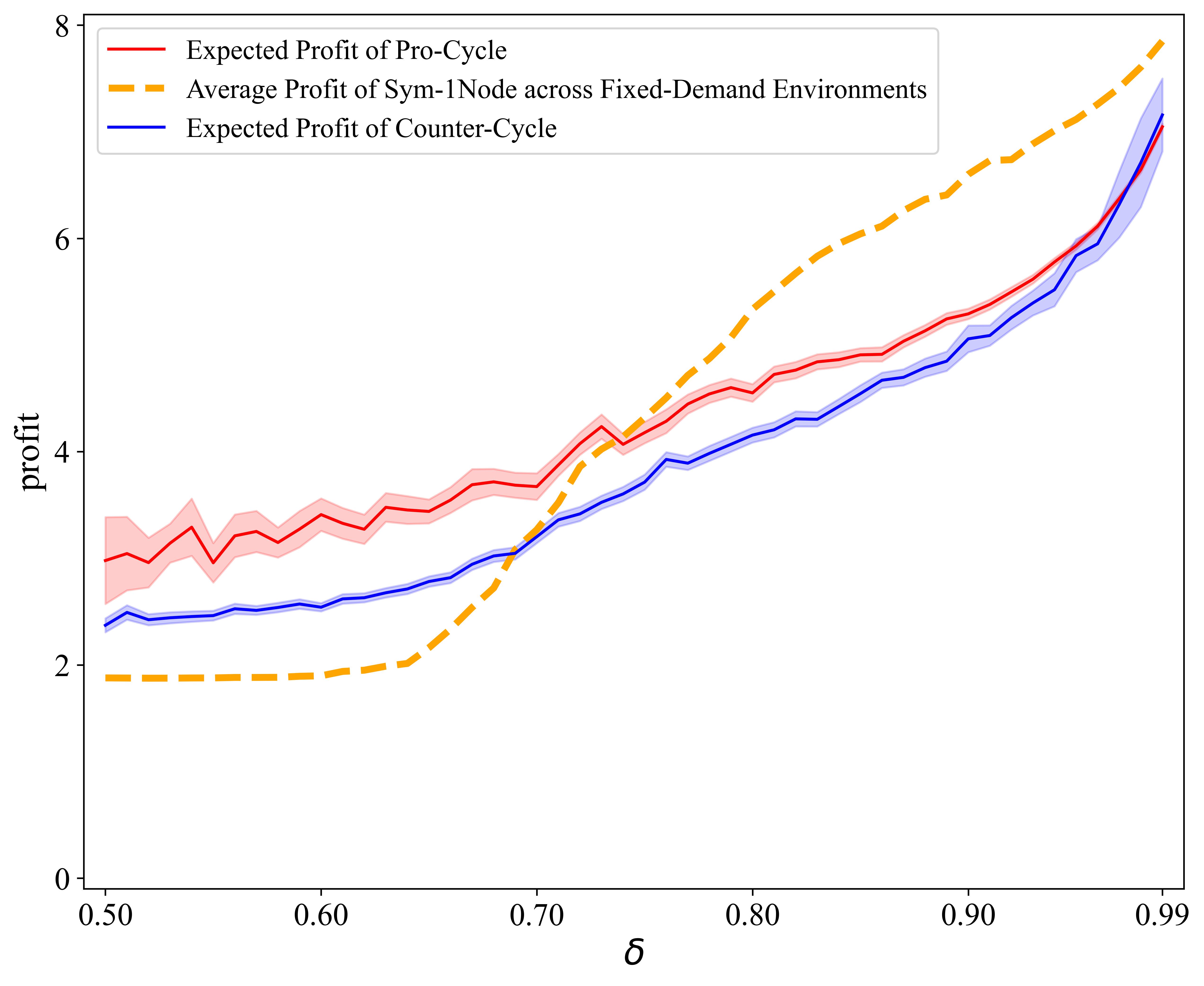}
\caption{Profit Dynamics across $\delta$}
\caption*{Notes: The red and blue lines plot the expected profits of Pro-Cycle and Counter-Cycle across $\delta$, with shaded areas indicating $5\%$ confidence intervals. The orange dashed line reports the average profit of Sym-1Node across two fixed-demand environments ($L$ and $H$), serving as a benchmark for evaluating demand-contingent pricing.}
\label{fig:profit_dynamics}
\end{figure}

The sharply contrasting profit outcomes at $\delta = 0.96$ and $\delta = 0.66$ raise a natural question: does the profit reversal between these two learning environments reflect a systematic pattern or a mere coincidence?
Figure \ref{fig:profit_dynamics} plots the expected profits of Counter-Cycle and Pro-Cycle across $\delta$, along with the average profit of Sym-1Node from the fixed-demand environments as a benchmark.
The figure shows a clear non-monotonic relationship: profits exceed the fixed-demand benchmark at low $\delta$ but fall below it once $\delta$ passes the turning points ($\delta=0.70$ for Counter-Cycle and $\delta=0.74$ for Pro-Cycle).

Thus, relative to fixed-demand learning, the profit reversal arises naturally from the demand-contingent pricing learned under observed demand shocks, highlighting that demand observability is a double-edged sword for profitability.

\begin{flushleft}
\textbf{Result 4:} 
Relative to learning under the fixed-demand benchmark, learning under observed demand shocks exhibits a profit reversal: it yields higher profits at low $\delta$, but lower profits at high $\delta$.
\end{flushleft}

\section{State Representations and Pricing Patterns} \label{sec:state_outcome}
Building on the preceding analysis, this section examines how the information contained in the state variable shapes the pricing patterns learned by Q-learning agents.

\subsection{Demand Memory}
Recall that the state variable is defined as $s_t=(\theta_{t-1}, p_{1t-1}, p_{2t-1}, \theta_t)$, incorporating the previous period’s demand state and price pair, together with the current demand realization.
Under i.i.d. demand shocks, observing $\theta_t$ is necessary for Q-learning agents to implement demand-contingent pricing.
A natural question then arises: is the lagged demand $\theta_{t-1}$ necessary?
It is theoretically redundant: monitoring deviations can only rely on the last period's price pair, and $\theta_{t-1}$ is payoff-irrelevant for future payoffs under the i.i.d. assumption on demand shocks.

However, whether this theoretical redundancy applies to Q-learning remains uncertain. To examine this, I conduct a simulation in which the state variable, referred to as \textit{No Demand Memory}, is simplified by removing $\theta_{t-1}$. Table \ref{table:state_partial} shows that, under this reduced specification, both demand-contingent pricing patterns (Pro-Cycle at $\delta=0.96$ and Counter-Cycle at $\delta=0.66$) remain prevalent but decline substantially relative to the full state representation. At the same time, Sym-Rigid increases significantly.\footnote{The proportion test indicates that all observed differences are statistically significant at the $p < 0.001$ level.} Thus, removing the previous demand state reduces the efficiency with which Q-learning agents learn demand-contingent pricing.

\begin{table}[h!]
\centering
\scalebox{0.99}{
\begin{threeparttable}
\caption{Frequencies of Pricing Patterns under Different State Representations} 
\label{table:state_partial}
\begin{tabular}{ccccc}
\hline
State   Variable & Pro-Cycle & Counter-Cycle & Sym-Rigid & Others \\ \hline
\multicolumn{5}{l}{\textit{Panel A: $\delta$=0.96}}                           \\
Full Memory\textsuperscript{a}      & 0.788     & 0.03          & 0.002     & 0.180  \\
\multicolumn{1}{l}{Reduced Settings:} &           &               &           &        \\
No Demand Memory\textsuperscript{b} & 0.448     & 0.071         & 0.278     & 0.203  \\
No Price Memory\textsuperscript{c}  & 0.788     & 0.108         & 0.104     & 0      \\
No Memory\textsuperscript{d}        & 0.797     & 0.125         & 0.002     & 0.076  \\
                 &           &               &           &        \\
\multicolumn{5}{l}{\textit{Panel B: $\delta$=0.66}}                           \\
Full Memory      & 0.165     & 0.600         & 0.033     & 0.202  \\
\multicolumn{1}{l}{Reduced Settings:} &           &               &           &        \\
No Demand Memory & 0.007     & 0.517         & 0.421     & 0.055  \\
No Price Memory  & 0         & 0.003         & 0.997     & 0      \\
No Memory        & 0         & 0             & 1         & 0      \\ \hline
\end{tabular}
\begin{tablenotes}
  \small
\item Notes: a. Full memory: $s_t = (\theta_{t-1}, p_{1t-1}, p_{2t-1}, \theta_t)$.
    \item b. No Demand Memory: $s_t = (p_{1t-1}, p_{2t-1}, \theta_t)$.
    \item c. No Price Memory: $s_t = (\theta_{t-1},\theta_t)$.
    \item d. No Memory: $s_t = (\theta_t)$.
\end{tablenotes}
\end{threeparttable}} 
\end{table}

One likely reason is that, without $\theta_{t-1}$, Q-learning agents can no longer distinguish whether a rival’s price change reflects a demand shock or a deviation. Confronted with this ambiguity, the learning process sometimes converges to rigid pricing that equalizes prices across demand states to eliminate uncertainty. Nonetheless, since the current demand $\theta_t$ remains observable, agents can still form a coarse mapping between demand states and payoffs, so demand-contingent pricing is weakened rather than eliminated.

\subsection{Price Memory}
I next examine the role of price memory by considering two reduced settings: \textit{No Price Memory} and \textit{No Memory}.
In the former, agents retain only demand information, with the state variable defined as $s_t = (\theta_{t-1}, \theta_t)$; in the latter, agents know only the current demand state, $s_t = (\theta_t)$.
Table \ref{table:state_partial} shows that, at $\delta = 0.96$, both settings learn Pro-Cycle, with frequencies nearly identical to those observed under the full state representation.
In contrast, when $\delta = 0.66$, Counter-Cycle cannot be learned under these two settings; instead, both agents converge to rigid pricing and only earn profits at the competitive level.

\begin{flushleft}
\textbf{Result 5:} 
Including demand memory in the state variable substantially reduces rigid pricing and strengthens the dominance of demand-contingent pricing. 
The role of price memory differs across $\delta$: at high $\delta$, observing current demand is sufficient for learning procyclical pricing regardless of whether price memory is present, whereas sustaining countercyclical pricing at low $\delta$ requires price memory.
\end{flushleft}

\subsection{Summary}
Based on these findings, I next explain how Q-learning agents generate distinctive demand-contingent pricing patterns across $\delta$.
Recall that in \cite{rs1986}, the emergence of the cyclicality of pricing patterns relies on two core insights: (i) the deviation incentive is always stronger during booms; and (ii) the trade-off between immediate rewards and continuation values is governed by $\delta$.

When $\delta$ is low, immediate gains dominate the learning process. Through exploration, Q-learning agents discover that undercutting the rival under $H$ yields higher Q-values for lower-price actions than under $L$, as immediate deviation gains are typically much larger under $H$.\footnote{Q-updates also include realized discounted continuation values. However, these continuation-value adjustments are similar on average for deviations originating from both demand states. As demand shocks are i.i.d. and Q-learning operates with bounded memory, deviations from $H$ and $L$ lead to indistinguishable post-deviation price dynamics. As a result, agents cannot learn harsher punishments for deviations under $H$.}
Since continuation values matter little when $\delta$ is low, charging the same price cannot sustain Q-values as high as those obtained from undercutting. As a result, agents are inclined to undercut each other under $H$, driving prices toward the competitive level. By contrast, undercutting under $L$ is less attractive, allowing agents to maintain relatively higher prices under $L$. This gives rise to countercyclical pricing.

When $\delta$ is high, continuation values play a dominant role in Q-updates, making high prices easier to sustain in both demand states. Through exploration, agents learn to charge higher prices under $H$, because doing so generates higher Q-values. Although deviation incentives remain stronger under $H$, the greater weight placed on continuation values offsets the immediate gains from undercutting, making coordinated high prices more resistant to short-run deviations. This generates procyclical pricing.

In this sense, even though Q-learning agents have no explicit knowledge of these two core insights, their Q-values encode them through pure trial and error. As a result, agents come to behave as if guided by these insights, thereby reproducing the pricing patterns predicted by collusion theory.

%%%%%%%%%%%%%%%%%%%%%%%%%%%%%%%%%%%%%%%%%%%%%%%%%%%%%%%%%%%%%%%%%%%%%%%%%%%%%%%%%%%%%%%%%%%%%%%%%%

%%%%%%%%%%%%%%%%%%%%%%%%%%%%%%%%%%%%%%%%%%%%%%%%%%%%%%%%%%%%%%%%%%%%%%%%%%%%%%%%%%%%%%%%%%%%%%%%%%
\section{Robustness Checks} \label{sec:robust}
This section examines the robustness of demand-contingent pricing and the associated supracompetitive profits. It also conducts deviation tests, and investigates how the information structure, particularly whether demand shocks are observed by both agents or by only one, shapes these learning outcomes.

\subsection{Alternative Initialization}
In the baseline setting, the Q-matrix is initialized based on the assumption of completely random play by the opponent. As a robustness check, I instead initialize all Q-values to zero, so that the agent begins without any prior preference over actions. The resulting pricing patterns remain consistent with the baseline. Specifically, Pro-Cycle occurs in $67.5\%$ of the sessions at $\delta = 0.96$, while Counter-Cycle appears in $64.7\%$ at $\delta = 0.66$.

\subsection{Hyperparameter Variations}
To evaluate how hyperparameters influence learning outcomes, I systematically vary the learning rate $\alpha$ and the experimentation parameter $\beta$. For each parameter, I construct a grid of ten equally spaced points: $\alpha \in [0.05, 0.5]$ and $\beta \in [10^{-6}, 10^{-5}]$.
The range of $\alpha$ captures different trade-offs between incorporating new information and retaining past experience, while the range of $\beta$ reflects varying degrees of exploration during the learning process (lower values of $\beta$ correspond to more extensive exploration).

\begin{figure}[h!]
\centering
\begin{subfigure}{0.32\textwidth}
    \includegraphics[width=\textwidth]{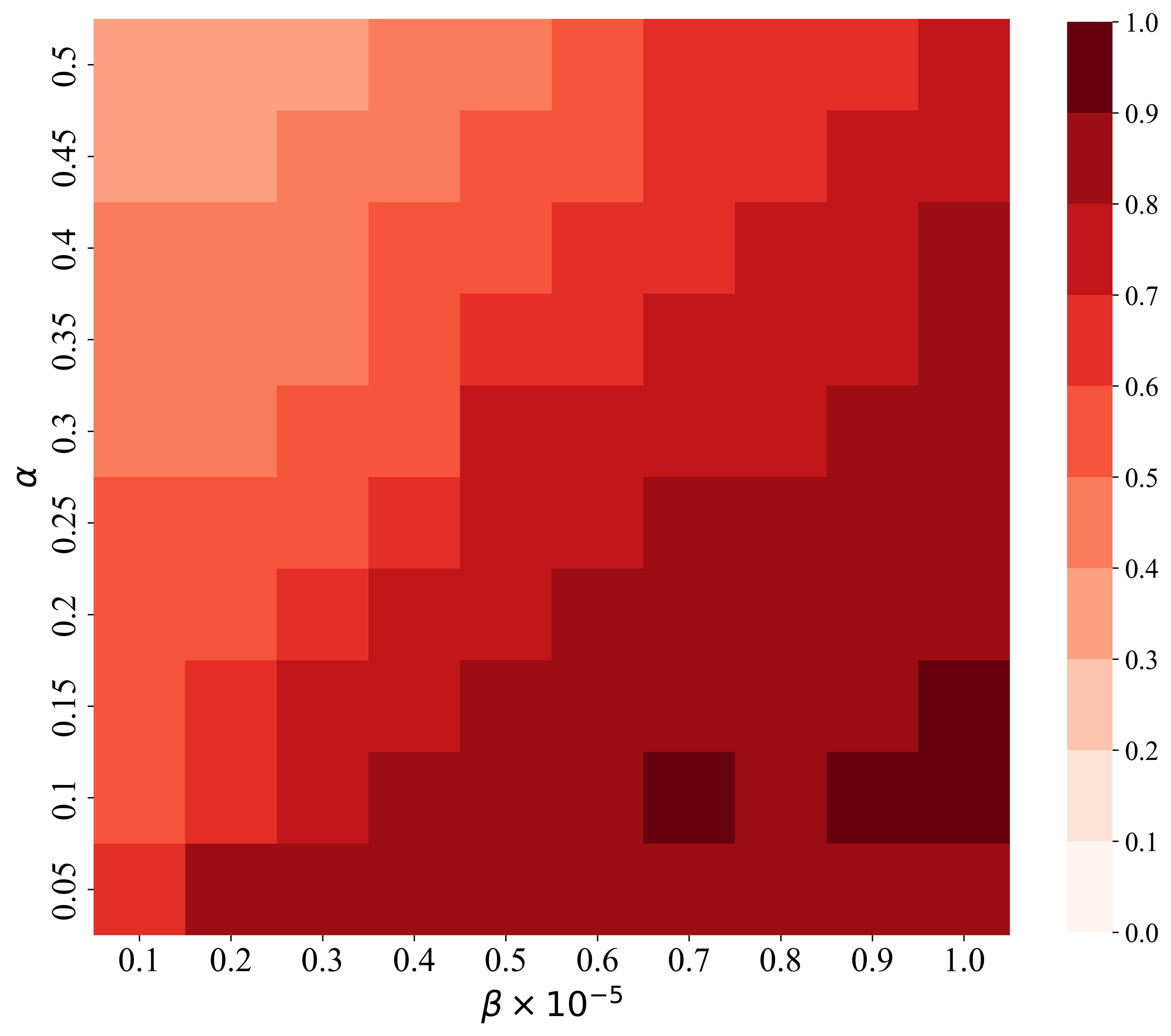}
    \caption{Frequency of Pro-Cycle}
    \label{fig:robust_96_1}
\end{subfigure}
  \hspace{2em}
\begin{subfigure}{0.32\textwidth}
    \includegraphics[width=\textwidth]{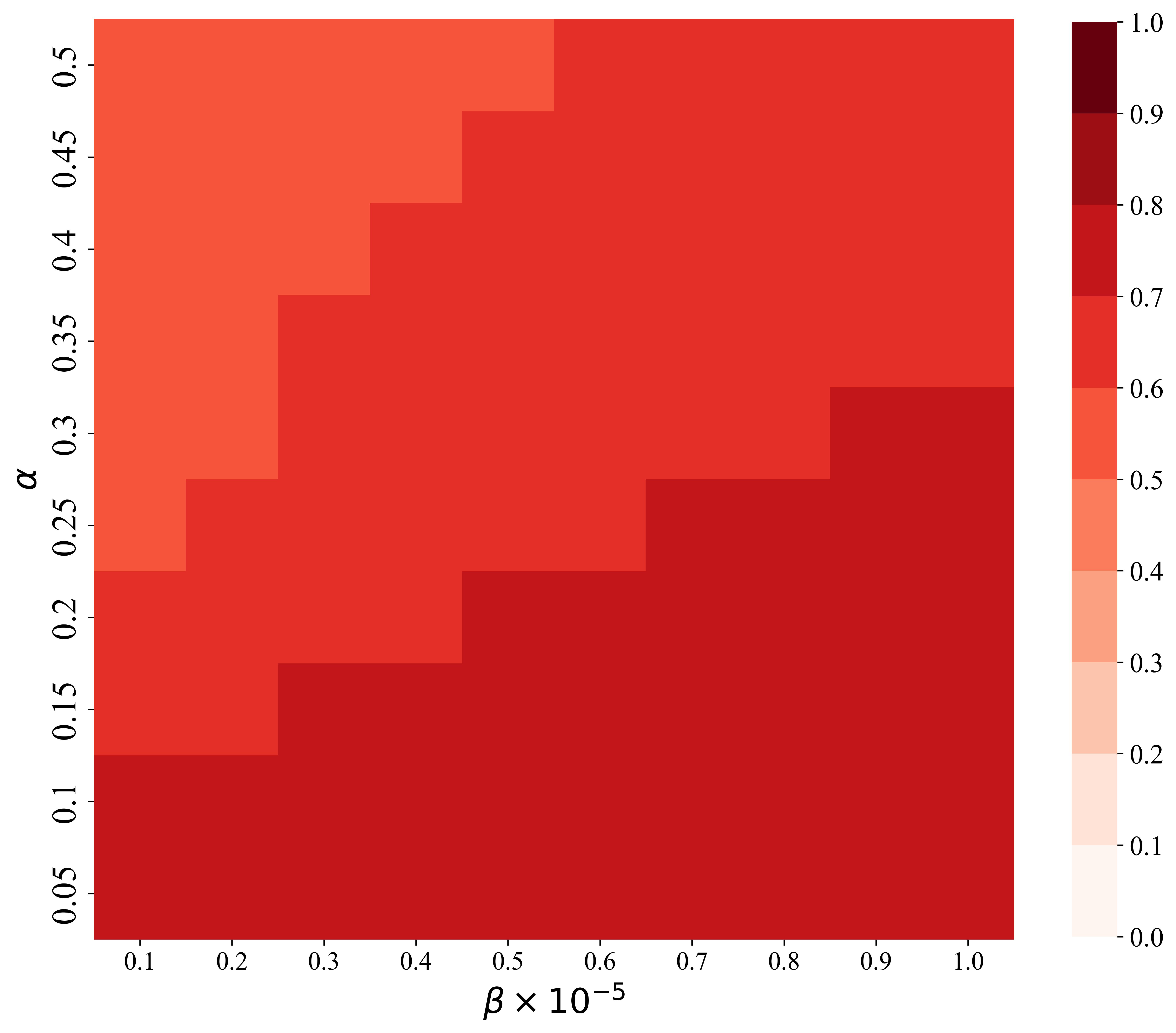}
    \caption{Profitability of Pro-Cycle}
    \label{fig:robust_96_2}
\end{subfigure}
\caption{Summary of Pro-Cycle across Parameter Space at $\delta=0.96$}
\label{fig:robust_96}
\end{figure}

\paragraph{Pricing Patterns}
Figure \ref{fig:robust_96_1} shows that, across a wide range of $\alpha$ and $\beta$, Pro-Cycle continues to be the predominant pricing pattern at $\delta=0.96$. It exceeds $50\%$ over most of the parameter space, and its prevalence increases as $\alpha$ decreases and $\beta$ increases. These results demonstrate that Pro-Cycle remains robustly dominant at high $\delta$.

\begin{figure}[h!]
\centering
\begin{subfigure}{0.32\textwidth}
    \includegraphics[width=\textwidth]{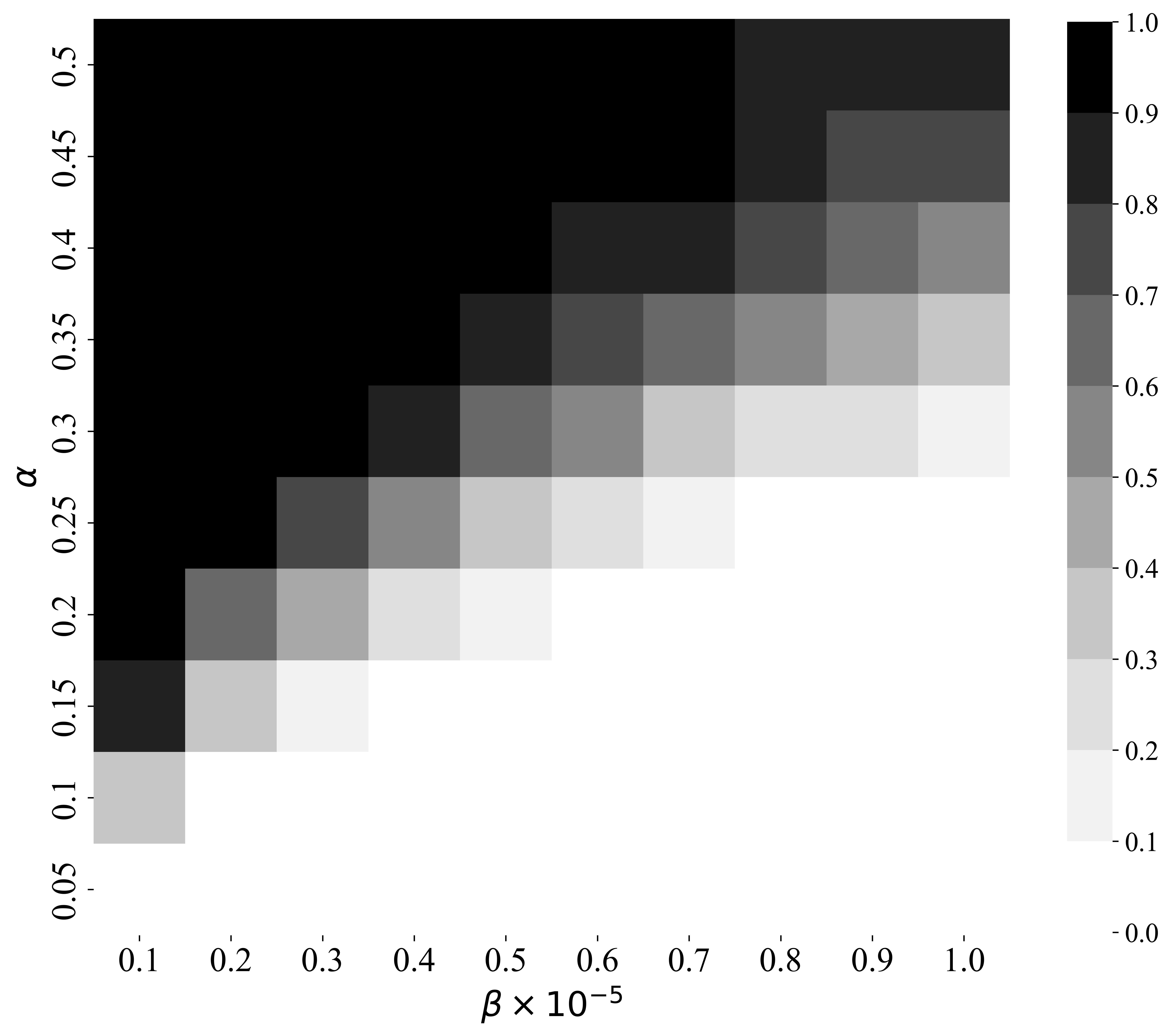}
    \caption{Sym-Rigid}
    \label{fig:robust_66_1}
\end{subfigure}
\hfill
\begin{subfigure}{0.32\textwidth}
    \includegraphics[width=\textwidth]{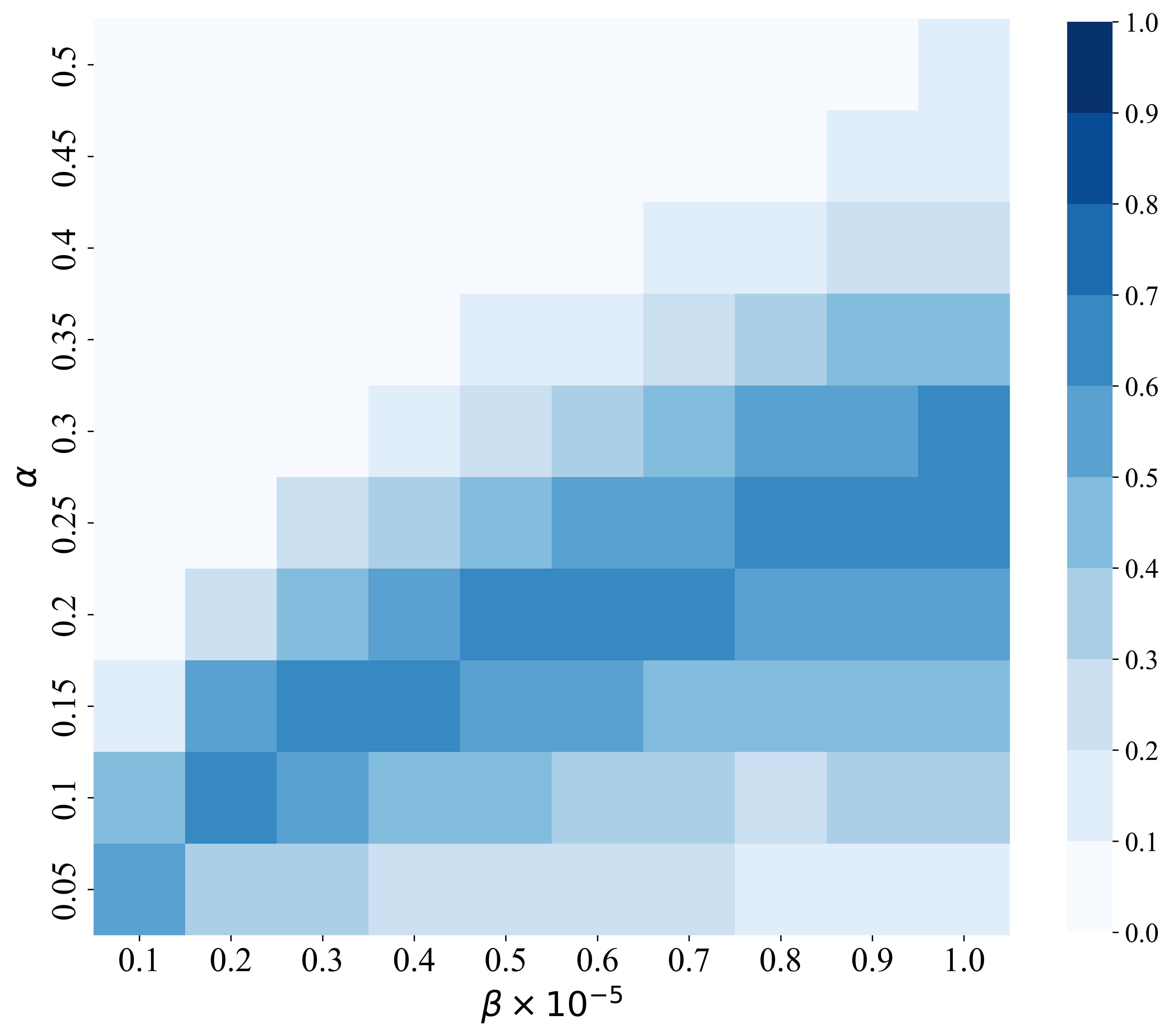}
    \caption{Counter-Cycle}
    \label{fig:robust_66_2}
\end{subfigure}
\hfill
\begin{subfigure}{0.32\textwidth}
    \includegraphics[width=\textwidth]{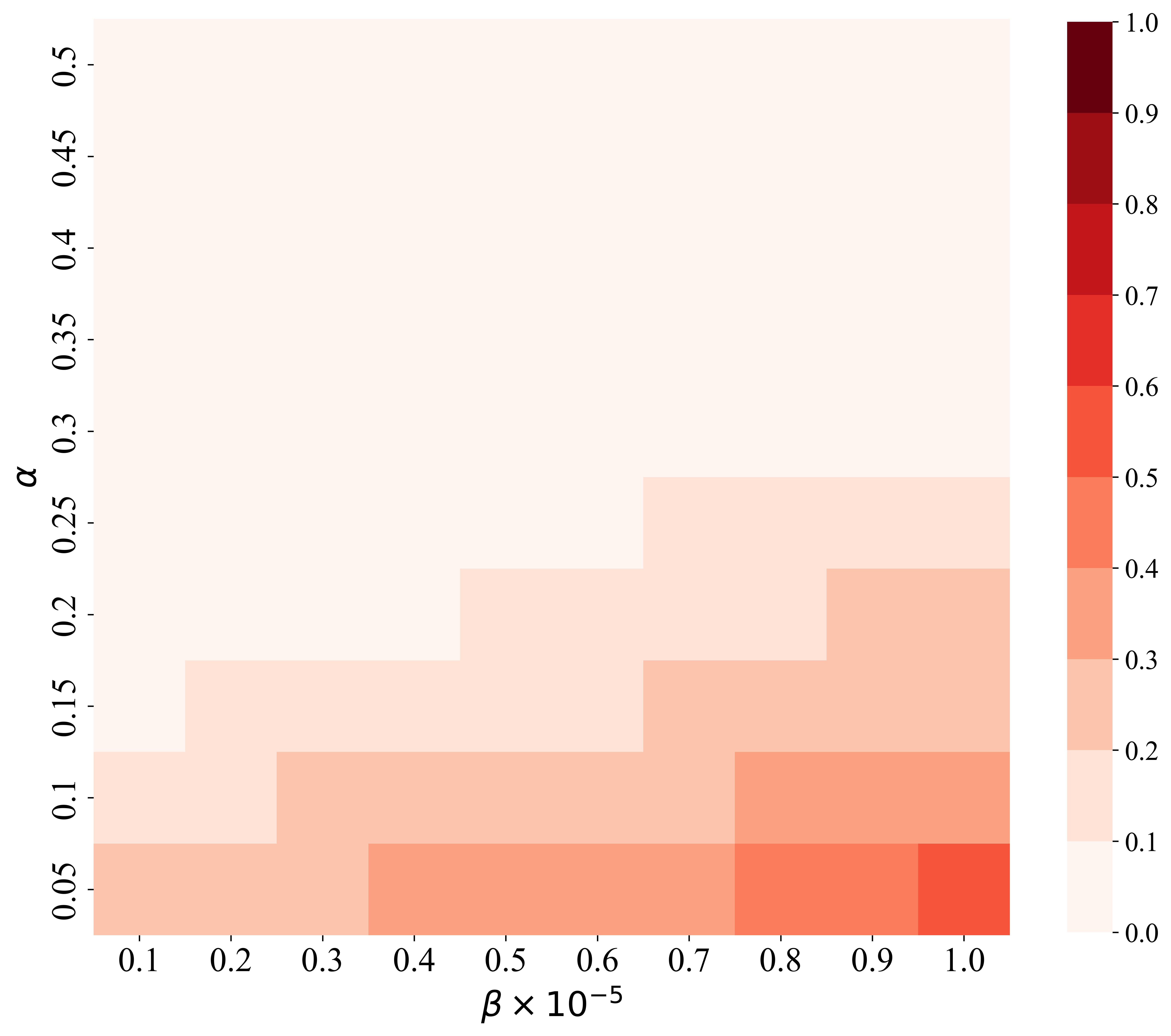}
    \caption{Pro-Cycle}
    \label{fig:robust_66_3}
\end{subfigure}
\caption{Frequencies of Different Pricing Patterns across Parameter Space at $\delta=0.66$}
\label{fig:robust_66}
\end{figure}

In contrast, Figure \ref{fig:robust_66} reveals substantial variation in which pricing pattern dominates when $\delta = 0.66$. Sym-Rigid dominates in the upper region of the hyperparameter space, whereas demand-contingent pricing patterns prevail in the lower region, among which Counter-Cycle dominates most of the area.
Thus, the prevalence of Counter-Cycle depends on the joint effect of the discount factor and the hyperparameters ($\alpha$ and $\beta$) that govern learning dynamics.

\paragraph{Supracompetitive Profits}
Figure \ref{fig:robust_96_2} presents the expected profits under Pro-Cycle ($\delta = 0.96$) as a function of $\alpha$ and $\beta$. Profits range from $50\%$ to $76\%$ of the best collusive outcome across the hyperparameter space. These results indicate that Q-learning agents can autonomously sustain supracompetitive profits under Pro-Cycle across a broad range of hyperparameters, without relying on artificial parameter choices.

\subsection{Deviation Tests}
I next evaluate how the reward-punishment scheme functions in sustaining supracompetitive profits under observed demand shocks.\footnote{According to \citet{abada2025algorithmic}, the supracompetitive profits observed in the baseline can be interpreted as arising from a reward-punishment scheme, given that price memory is included. See that paper for further discussion.}
Specifically, I ask within the price cycle, if one agent undertakes its most profitable deviation at the current node, is that deviation profitable? Focusing on the most profitable deviation establishes a lower bound for evaluation: if even the most tempting deviation is unprofitable, all less aggressive or randomly explored deviations will be unprofitable as well.\footnote{The deviation test follows \citet{calvano2020}, and the modifications, along with the deviation rule and procedure, are described in the Appendix \ref{app:deviation}.}

Figure \ref{fig:unprofitable_rate} in the Appendix presents the probabilities of unprofitable deviations across $\beta$ under Pro-Cycle. The parameter $\beta$ governs exploration intensity, with $\alpha = 0.15$ and $\delta = 0.96$ held fixed.
The probability conditional on $L$ is higher than that on $H$, reflecting lower short-term deviation gains at low demand. Overall, the likelihood of unprofitable deviations stays near $50\%$ across most $\beta$ and rises above $60\%$ only when exploration becomes extremely extensive. 
Given the random exploration inherent in Q-learning, the fact that roughly half of the most profitable deviations are unprofitable suggests that the reward-punishment scheme is strong enough to sustain supracompetitive profits.

\subsection{Asymmetric Information about Demand Shocks}
I finally examine how asymmetric information about demand shocks affects learning outcomes.\footnote{For reference, \citet{bm2023} study the asymmetry in pricing algorithms that arises from differences in updating frequency.} In practice, firms may deploy pricing algorithms with differing access to demand information, raising the question of whether Q-learning agents facing such informational asymmetries can still coordinate and reproduce the baseline learning outcomes. Specifically, I consider a setting in which one firm observes the demand state and incorporates it into its pricing algorithm, while the other does not.

\subsubsection{Pricing Patterns}
\begin{figure}[h!]
\centering
\begin{subfigure}{0.58\textwidth}
    \includegraphics[width=\textwidth]{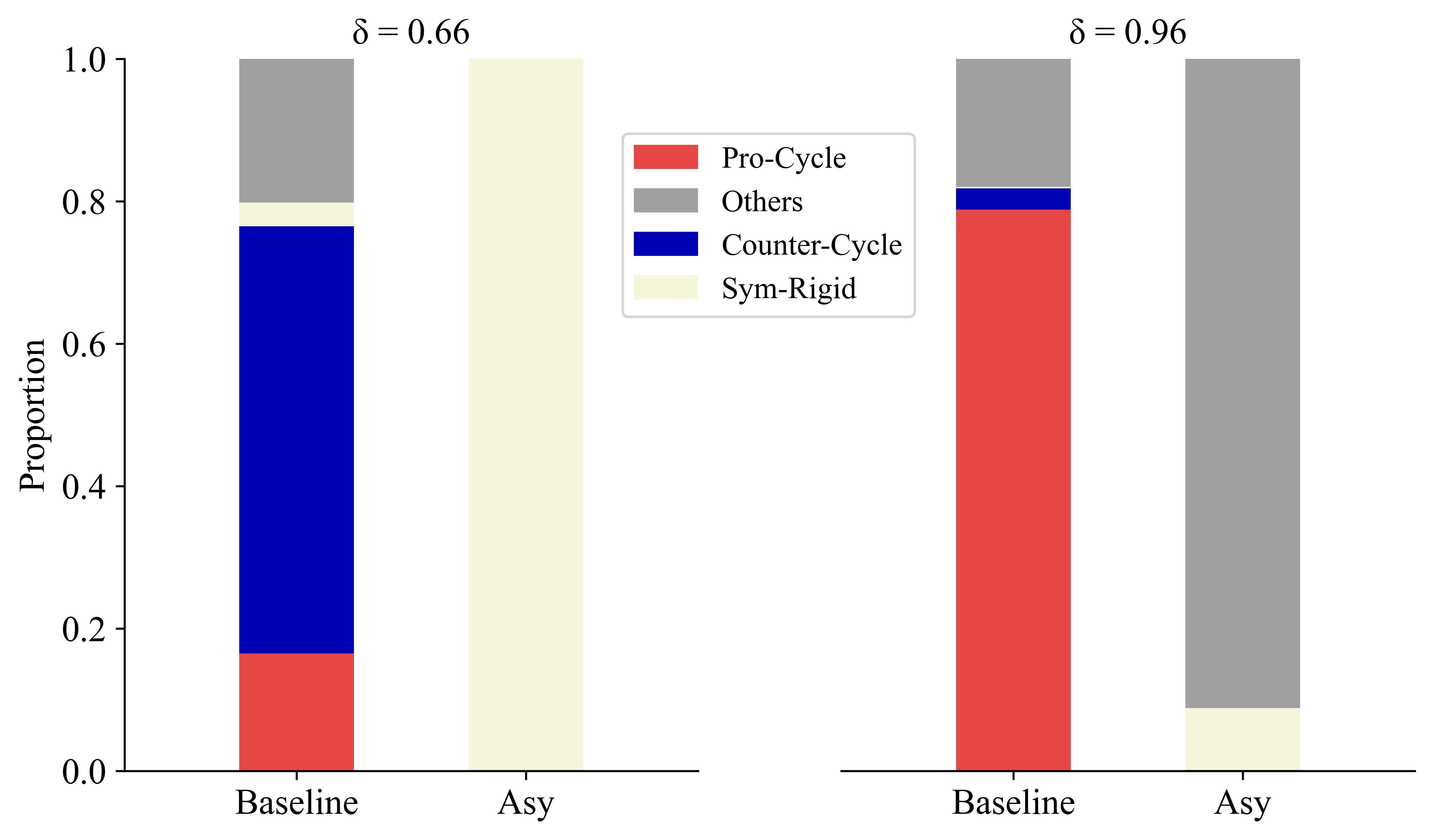}
    \caption{Distribution of Pricing Patterns}
    \label{fig:asy_price}
\end{subfigure}
\hfill
\begin{subfigure}{0.4\textwidth}
    \includegraphics[width=\textwidth]{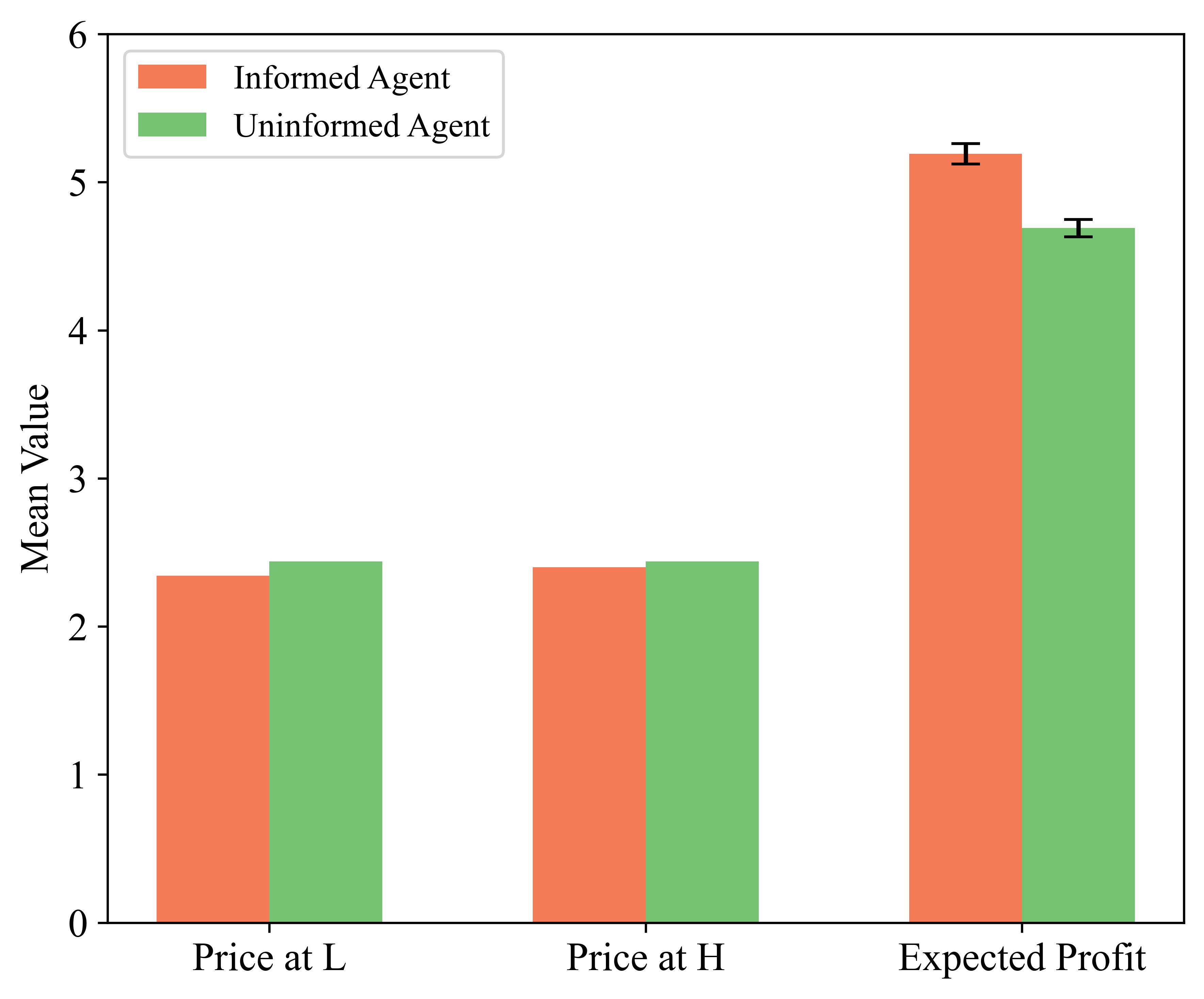}
    \caption{Informed vs. Uninformed}
    \label{fig:asy_profit}
\end{subfigure}
\caption{Learning Outcomes under Asymmetric Demand Information}
\caption*{Notes: Figure \ref{fig:asy_price} compares pricing patterns under asymmetric and symmetric information (baseline).
Figure \ref{fig:asy_profit} compares the informed and uninformed agents’ prices and profits under the pricing pattern of Semi-Rigid, where the uninformed agent keeps prices fixed across demand states, while the informed agent varies them with the demand state.
}
\label{fig:asymmetry}
\end{figure}

Figure \ref{fig:asy_price} compares the distribution of pricing patterns under asymmetric and symmetric information (baseline) at $\delta = 0.66$ and $\delta = 0.96$. The figure shows that once demand observability becomes asymmetric, coordination in demand-contingent pricing breaks down completely, eliminating both Pro-Cycle and Counter-Cycle.
This finding underscores that demand-contingent pricing can be sustained only when both Q-learning agents observe the demand state and interpret price movements in the same way.

Instead, Sym-Rigid for the informed agent (and Sym-1Node for the uninformed agent) reaches $100\%$ at $\delta = 0.66$, where prices are uniformly fixed at $0.5$. The mechanism is straightforward: when only one agent observes demand shocks and adjusts prices accordingly, the uninformed agent cannot tell whether a price change reflects coordination or deviation. To avoid being undercut when immediate rewards are weighted more heavily, the uninformed agent sets a low price, which the informed agent follows, triggering a price war that drives prices down to the competitive level.

At $\delta = 0.96$, however, Sym-Rigid constitutes only $8.7\%$ of the sessions. Instead, about $91.2\%$ of the sessions are classified as Others. Among them, almost all converge to a pricing pattern referred to as \textit{Semi-Rigid}, where the uninformed agent keeps prices fixed across demand states, while the informed agent varies them with the demand state. Figure \ref{fig:asy_profit} shows that the informed agent earns, on average, $10.7\%$ more in profits than the uninformed agent (paired t-test, $p<0.001$). This suggests that at high $\delta$, greater patience allows the uninformed agent to tolerate occasional losses from being undercut while splitting the market evenly most of the time, thereby enabling the informed agent to extract an information premium.

\section{Conclusion} \label{sec:conclusion}
This paper examines how the observability of demand shocks shapes the learning outcomes of pricing algorithms. The simulation results show that Q-learning agents autonomously adapt to demand fluctuations and develop demand-contingent pricing, which is procyclical at high $\delta$ and countercyclical at low $\delta$, consistent with \citet{rs1986}. They also sustain supracompetitive profits, demonstrating the robustness of algorithmic collusion under observed demand shocks. 
Further analysis reveals that the learned pricing patterns critically depend on the information embedded in the state variable. Removing demand memory weakens demand-contingent pricing and drives learning toward rigid pricing. Price memory has a more complex effect: at high $\delta$, agents learn procyclical pricing regardless of whether price memory is present, whereas at low $\delta$, the absence of price memory leads to uniformly rigid pricing, making it essential for sustaining countercyclical pricing.
These findings shed light on how Q-learning agents generate distinctive demand-contingent pricing patterns across $\delta$. Through repeated exposure to payoff variation across demand states and prices, they come to internalize both the stronger deviation incentives that arise during booms and the discount factor’s role in balancing short-term gains and long-term continuation values, thereby reproducing the pricing patterns predicted by collusion theory.

% direction 1
Although this paper advances our understanding of algorithmic pricing under i.i.d. observed demand shocks, a broader question remains: would Q-learning algorithms continue to learn and behave in line with theoretical predictions when the demand environment changes? Future work could address this question by relaxing the i.i.d. assumption and examining demand structures that exhibit serial correlation \citep{k1991correlated}, persistence over time \citep{bagwell1997collusion}, deterministic business-cycle patterns \citep{hh1991cycle}, or firm-specific shocks. Comparing theoretical predictions with actual learning outcomes in such environments would reveal whether Q-learning algorithms can consistently learn the outcomes predicted by theory, thereby shedding light on whether they truly capture the underlying economic mechanisms.

% direction 2
This paper also uncovers a profit reversal: learning under observed demand shocks yields higher profits than the fixed-demand benchmark when $\delta$ is low but lower profits when $\delta$ is high. This highlights that demand observability is a double-edged sword for profitability. A promising direction for future research is developing a formal framework that characterizes the learning dynamics underlying this reversal and exploring how it extends to richer information environments or more complex market structures.

\section*{Acknowledgement}
Computation reported in this work was carried out on the Unity Cluster of the College of Arts and Sciences at the Ohio State University. The computational resources and support provided are gratefully acknowledged.

\newpage
\appendix
\section*{Appendix}

\renewcommand{\thesubsection}{\Alph{subsection}}
\renewcommand{\thefigure}{A.\arabic{figure}}
\setcounter{figure}{0}
\renewcommand{\thetable}{A.\arabic{table}}
\setcounter{table}{0}

% \subsection{Tables}

\subsection{Figures}

\begin{figure}[H]
\centering
\begin{subfigure}{0.35\textwidth}
    \includegraphics[width=\textwidth]{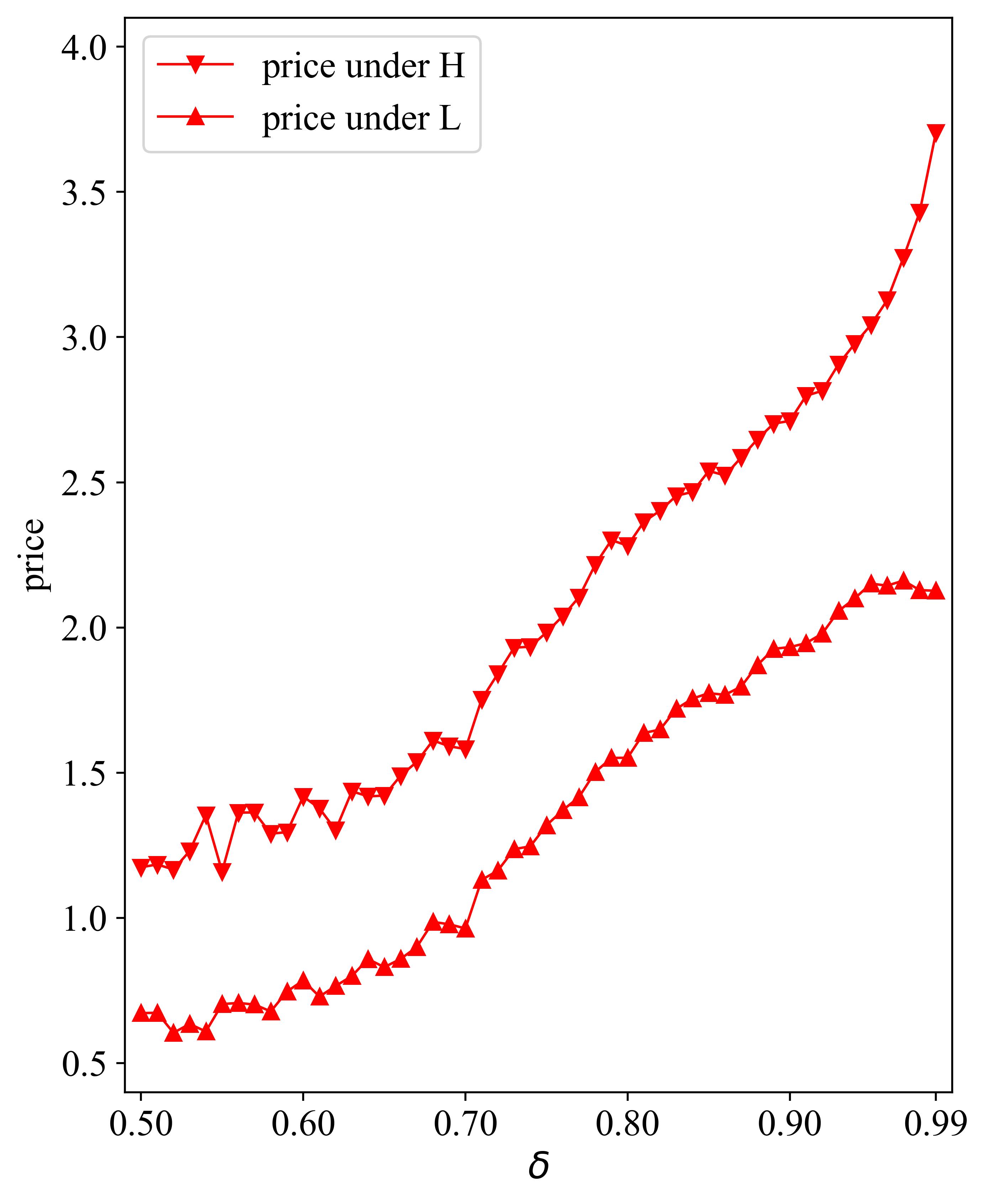}
    \caption{Pro-Cycle}
    \label{fig:price_pc}
\end{subfigure}
% \hfill
\hspace{0.05\textwidth}
\begin{subfigure}{0.35\textwidth}
    \includegraphics[width=\textwidth]{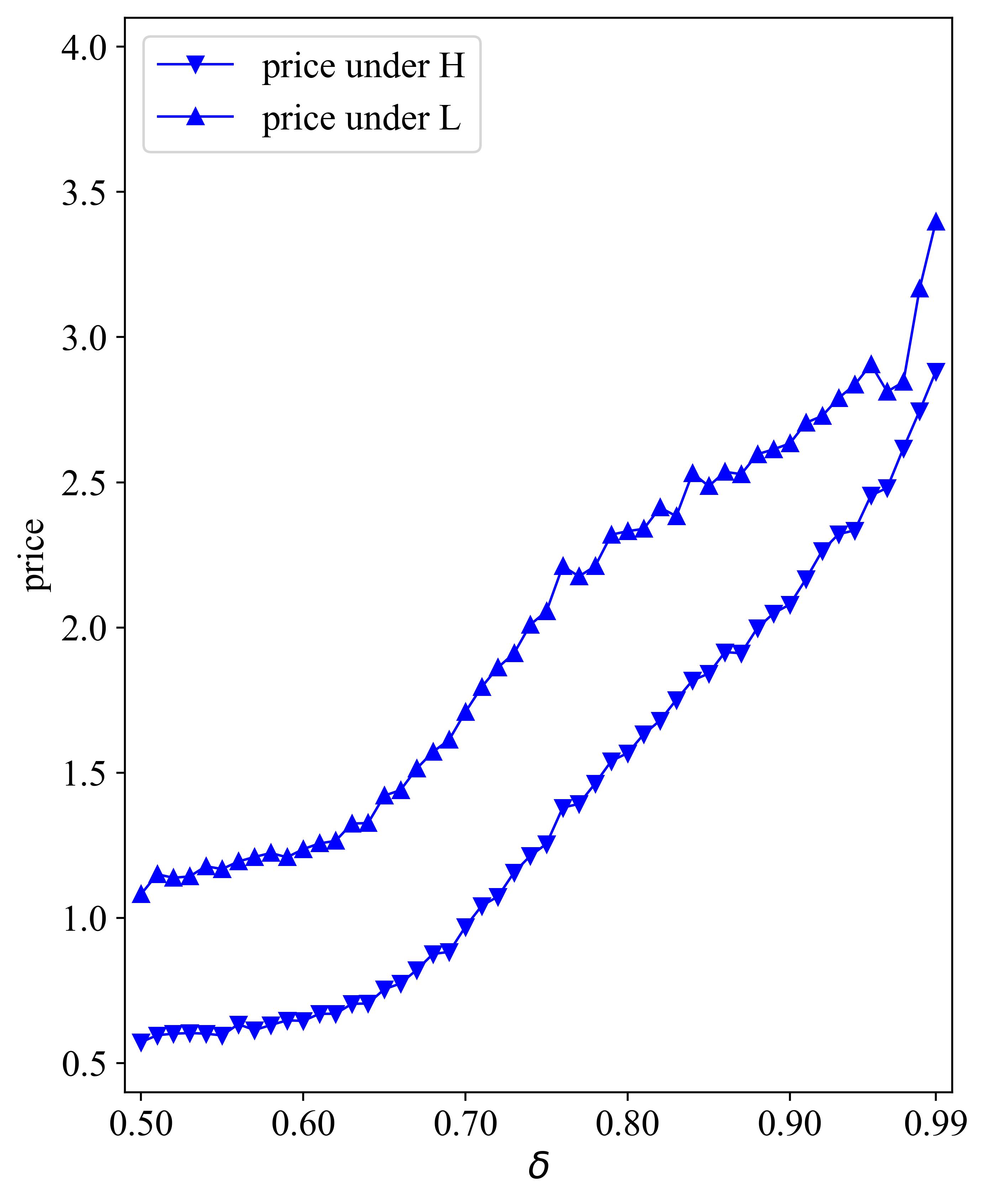}
    \caption{Counter-Cycle}
    \label{fig:price_cc}
\end{subfigure}
\caption{Price Dynamics of Demand-contingent Pricing}
\label{fig:price_pattern_delta}
\end{figure}

\begin{figure}[H]
\centering
\includegraphics[width=0.7\textwidth]{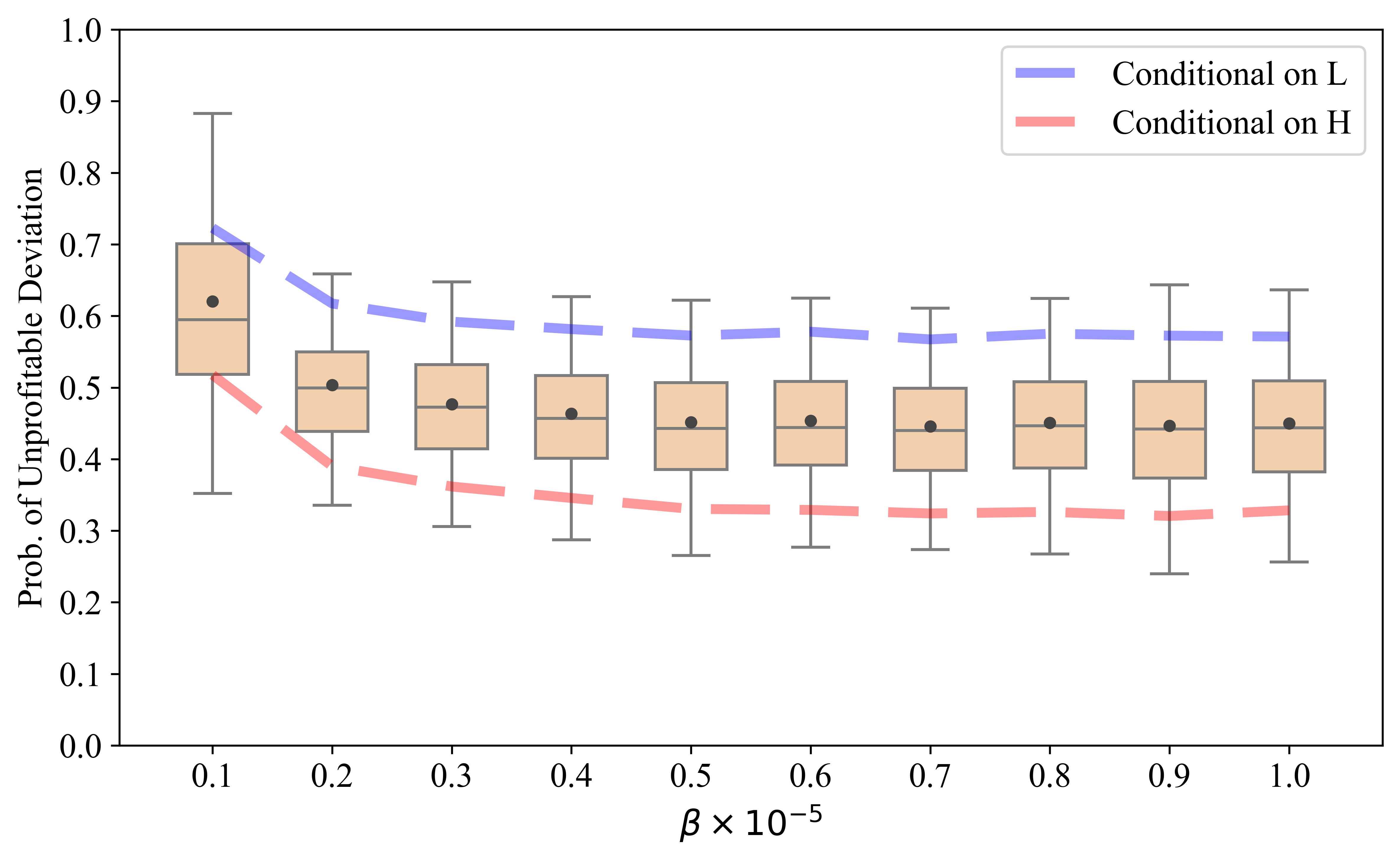}
\caption{Unprofitable Deviation vs. Exploration Rate}
\caption*{Notes: The box plot shows the probabilities of unprofitable deviation under Pro-Cycle, based on the most profitable deviation rule, with $\alpha = 0.15$ and $\delta = 0.96$ held fixed. Whiskers denote $1\times\mathrm{IQR}$, black dots means, and dashed lines conditional means by demand state (blue for $L$, red for $H$).}
\label{fig:unprofitable_rate}
\end{figure}

\begin{figure}[H]
\centering
\begin{subfigure}[b]{\textwidth}
    \centering
    \includegraphics[width=0.7\textwidth]{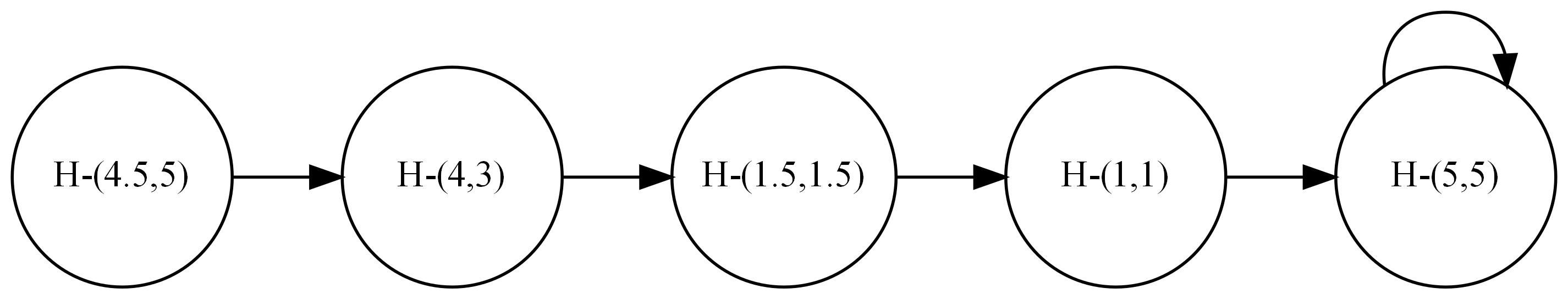}
    \caption{Single-path Deviation under Fixed Demand $H$}
    \label{fig:deviation_path_1}
\end{subfigure}

\vspace*{0.5cm}

\begin{subfigure}[b]{\textwidth}
    \centering
 \makebox[\textwidth][c]{\includegraphics[width=0.95\paperwidth]{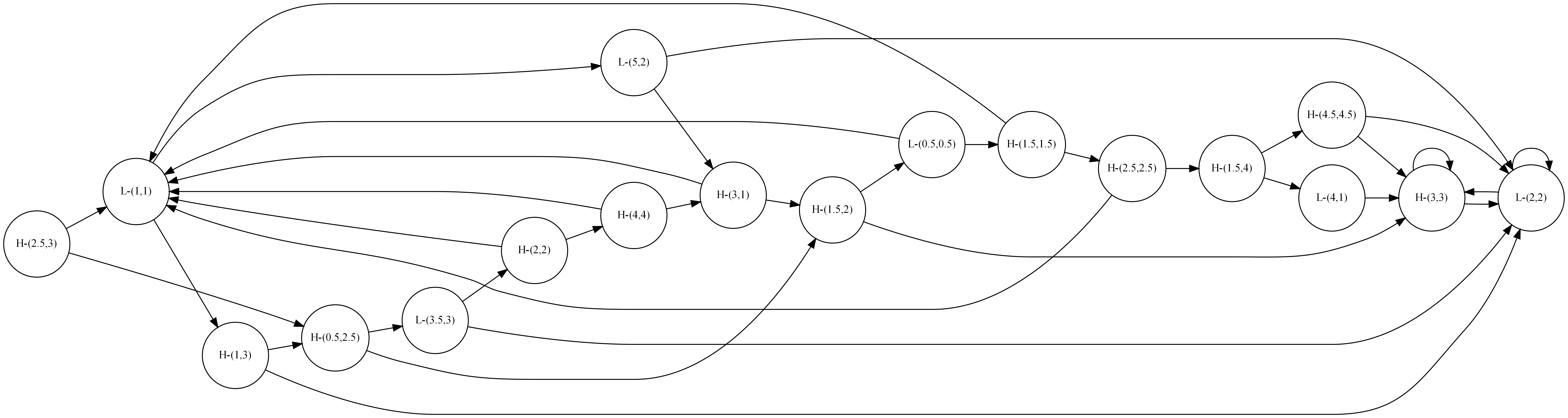}}
    \caption{Multi-path Deviation under Observed Demand Shocks}
    \label{fig:deviation_path_2}
\end{subfigure}
\caption{Impact of Observed Demand Shocks on Deviation Dynamics}
\caption*{Notes: Figure \ref{fig:deviation_path_1} illustrates a single deviation path under fixed demand $H$, where agent 1 undercuts the stable price by 0.5 and the price dynamics follow a deterministic path back to the price cycle. Figure \ref{fig:deviation_path_2} shows that once demand shocks are introduced, deviations generate multiple stochastic paths before the price dynamics return to the price cycle.}
\label{fig:deviation_path}
\end{figure}

\subsection{Initial Q-matrix}\label{app:initial} 
The initial Q-matrix $\mathbf{Q}_{i0}$ in the baseline setting is calculated as follows.
Given that the opponent randomly chooses a price at period $t=0$, the expected period payoff for agent $i$ who sets price $p_i$ in the demand state $\theta$ is
$$ 
\bar{\pi}_i(p_i,\theta)=\frac{ \sum_{p_{-i} \in A}\pi_i(p_i,p_{-i},\theta)}{|A|}
$$
Correspondingly, the initial Q-value at entry $(s_{\theta},p)$, where $s_{\theta}$ denotes any state with the current demand state being $\theta$, is
$$
Q_{i0}(s_{\theta},p_i)=Q_{i0}(\theta,p_i)=\bar{\pi}_i(p_{i},\theta) 
+ \delta \frac{\sum_{\theta_j \in \Theta}Q_{i0}(\theta_j,p_i)}{|\Theta|}
$$
where $Q_{i0}(s_{\theta},p_i)$ is written as $Q_{i0}(\theta,p_i)$ for simplicity. Thus, the expected continuation value equals the average Q-value of choosing $p_i$ across all possible demand states.

Then the initial Q-value at $p_i$ for each demand state can be solved through the linear equation system
$$
  \begin{array}{@{}*{5}{c@{}}}
    Q_{i0}(\theta_1,p_i)& {}={} &\bar{\pi}_i(p_{i},\theta_1) 
& {}+{} &\delta \frac{\sum_{\theta_j \in \Theta}Q_{i0}(\theta_j,p_i)}{|\Theta|}\\
    Q_{i0}(\theta_2,p_i)& {}={} &\bar{\pi}_i(p_{i},\theta_2) 
& {}+{} &\delta \frac{\sum_{\theta_j \in \Theta}Q_{i0}(\theta_j,p_i)}{|\Theta|}\\
    \vdots       &   & \vdots                &       &  \vdots                                       \\
    Q_{i0}(\theta_{|\Theta|},p_i)& {}={} &\bar{\pi}_i(p_{i},\theta_{|\Theta|}) 
& {}+{} &\delta \frac{\sum_{\theta_j \in \Theta}Q_{i0}(\theta_j,p_i)}{|\Theta|}
  \end{array} 
$$

\subsection{Average Long-run Prices}\label{app:steady}

\subsubsection{Directed Graph} 
Using the limit strategies $p_1^*(s)$ and $p_2^*(s)$, I construct a directed graph to represent the price dynamics. Consider a directed graph $G = (V, g)$, where $V = \{1, 2, \ldots, M\}$ is the set of nodes.\footnote{Note that $M = 242$.} Each node $v = (\theta, p_1, p_2)$ represents a combination of the demand state and prices charged by each agent.
The adjacency matrix $g$ is an $M \times M$ matrix with each element $g_{ij} \in \{0,1\}$ denoting whether a directed edge exists between nodes $i$ and $j$.\footnote{In a directed network, $g_{ij}$ generally does not equal $g_{ji}$.}
A directed edge exists (i.e., $g_{ij} = 1$) in $G$ if and only if the prices in node $v_j$ are optimal responses prescribed by the limit strategies, given the previous demand state and prices in node $v_i$ and the current demand state in node $v_j$. 
Note that it's possible to have $g_{ii} = 1$, which means that if the demand state remains the same from one period to the next, agents do not change their optimal prices, resulting in a self-loop.

\subsubsection{Price Cycle}
Price dynamics are expected to converge to a subgraph $G_c$, referred to as the price cycle, which is constructed to ensure that, once entered, all subsequent price movements remain permanently within it.
Formally, the price cycle $G_c$ must satisfy two requirements.
First, $G_c$ must be a strongly connected component (SCC), meaning that any node in $G_c$ can reach any other node through directed paths.\footnote{To identify SCCs in $G$, I employ the algorithm from \cite{tarjan1972}, as modified by \cite{nuutila1994}.}
Second, each node in $G_c$ must have all its direct successors within $G_c$. This ensures that random demand shocks cannot cause exits from $G_c$.\footnote{The number of outgoing edges for each node equals the number of possible demand states, which is two in the baseline.}
By construction, $G_c$ is an absorbing subgraph, as there exists no path from any node in $G_c$ to any node outside it.

The transitions on $G_c$ satisfy the Markov property, where each node has exactly two direct successors with equal transition probability of $0.5$. Therefore, this defines a finite Markov process. Since $G_c$ is strongly connected, the existence and uniqueness of its stationary distribution are guaranteed.
The stationary distribution $\psi^*$ (row vector) is defined by $\psi^* P = \psi^*$, where $P$ is the transition probability matrix (Markov matrix).\footnote{Every finite Markov model has at least one stationary distribution $\psi^*$. If the digraph is strongly connected, then $\psi^*$ is unique and everywhere positive. See this theorem in \cite{stokey1989recursive} and \cite{sargent2024economic}.}

Let $n$ denote the number of nodes in $G_c$. Let $\mathbb{1}_n$ denote the $1 \times n$ row vector $(1, \ldots, 1)$, $\mathbb{1}_{n \times n}$ the $n \times n$ matrix of ones, and $I$ the identity matrix.
With $P$ as the transition probability matrix of $G_c$, the unique stationary distribution $\psi^*$ is obtained by solving
\begin{equation}
\mathbb{1}_n = \psi^* (I-P+\mathbb{1}_{n \times n}) \label{markov_solve}
\end{equation}
Using the stationary distribution $\psi^*$, the average long-run price is then computed.
To illustrate, consider the transition probability matrix $P$ in Figure \ref{fig:exampleG_2}:
$$
\bordermatrix{      & L-(0.5,0.5)   & L-(2.5,2.5)   & H-(0.5,0.5) \cr
        L-(0.5,0.5)     & 0.5  & 0   & 0.5 \cr
        L-(2.5,2.5)     & 0.5  & 0   & 0.5 \cr
        H-(0.5,0.5)     & 0  & 0.5 & 0.5 \cr
                  }
$$
Solving equation \eqref{markov_solve} yields the stationary distribution $\psi^* = (0.25, 0.25, 0.5)$, where each element represents the steady-state probability of the corresponding node. The average long-run price pairs conditional on $L$ and $H$ are $(1.5, 1.5)$ and $(0.5,0.5)$, respectively.

% appendix
\subsection{Deviation Test}\label{app:deviation} 

\subsubsection{Modifications} 
The deviation test requires several modifications under observed demand shocks. 
First, with demand shocks, price dynamics form a finite Markov process in which each node is visited with a frequency determined by the stationary distribution. Therefore, when evaluating deviation over the entire price cycle, I take the weighted average using the stationary distribution.
Second, because i.i.d. demand shocks occur in each period, the deviation path is no longer deterministic but consists of multiple possible deviation paths.\footnote{Figure \ref{fig:deviation_path} in the Appendix illustrates this difference.} To address this randomness, I conduct one thousand repetitions for each node and take the average.
Third, asymmetric pricing may arise, so the deviation rule is adjusted accordingly.

\subsubsection{Deviation Rule}
The deviation test is implemented by applying the most profitable undercut.
The lowest feasible undercutting price is $\underline{p}=0.5$, as charging $0$ yields zero profit and is therefore excluded. 
In each period, the price pair is $(p_i,p_j)$, and the current demand state $\theta$ determines the monopoly price $p^M_\theta$. 
Given the opponent’s price $p_j$, the set of feasible undercutting prices for player $i$ is defined as 
$U(p_j) = \{ p  \mid p < p_j \}$.

The price that yields the most profitable deviation is denoted by $d_i(\theta,p_i,p_j)$.
The complete rule governing $d_i(\theta,p_i,p_j)$ is summarized in Table~\ref{table:deviation_rule}.
If $d_i=\varnothing$, the deviation is seen as unprofitable.
In Case 2, where $p_i>p_j$, $d_i$ is set to $0$ when $p_j=0$ to exogenously trigger an exit from the price cycle.

\begin{table}[h!]
\centering
\renewcommand{\arraystretch}{0.95}
\setlength{\tabcolsep}{4pt}
\small
\begin{tabular}{|c|c|c|}
\hline
\textbf{Case 1: } $p_i=p_j$ &
\textbf{Case 2: } $p_i>p_j$ &
\textbf{Case 3: } $p_i<p_j$ \\
\hline
$\begin{aligned}
&\begin{cases}
p^M_\theta & \text{if } p^M_\theta\in U(p_j)\\
p_i-0.5 & \text{if } p_i>0.5\\
\varnothing & \text{otherwise}
\end{cases}
\end{aligned}$

&

$\begin{aligned}
&\begin{cases}
p^M_\theta & \text{if } p^M_\theta\in U(p_j)\\
p_j-0.5 & \text{if } p_j>0.5\\
0.5 & \text{if } p_j=0.5\\
0 & \text{if } p_j=0
\end{cases}
\end{aligned}$

&

$\begin{aligned}
&\begin{cases}
p^M_\theta & \text{if } p^M_\theta\in U(p_j)\\
p_j-0.5 & \text{if } p_j-p_i>0.5\\
0.5 & \text{if } p_j=0.5\\
\varnothing & \text{otherwise}
\end{cases}
\end{aligned}$ \\
\hline
\end{tabular}
\caption{Deviation Rule under Three Price Relations}
\label{table:deviation_rule}
\end{table}

\subsubsection{Deviation Procedure}
The deviation test for a given node within the price cycle is conducted as follows. One agent is forced to deviate unilaterally from this node, after which both agents continue to follow their limit strategies until the price dynamics return to the price cycle, so that the sequence of prices becomes identical to that which would have occurred had no deviation taken place. The discounted profits obtained along this deviation path are then compared with the counterfactual profits that would have been realized along the non-deviation path within the price cycle to determine whether the deviation is profitable. Each node is simulated one thousand times to account for the stochasticity of demand shocks.

Algorithm~\ref{alg2} summarizes the procedure for conducting one deviation simulation. $\Pi^*_i$ and $\Pi^D_i$ denote agent $i$'s accumulated discounted profits on the non-deviation and deviation paths, respectively. The corresponding period profits are denoted by $\pi^*_{it}$ and $\pi^D_{it}$.

\begin{algorithm}[h!]
\caption{One Simulation for Deviation Test}\label{alg2}
\begin{algorithmic}[1]
% \Require Price dynamics is in $G_c$
\Statex \textit{\textbf{Step 1: Initialization}}

\State Deviation happens in the current node
\State $(\Pi^*_1, \Pi^*_2) \gets (\pi^*_{11}, \pi^*_{21})$ 
\Comment{on the non-deviation path}
\State Force one agent to undercut 
\State $(\Pi^D_1, \Pi^D_2) \gets (\pi^D_{11}, \pi^D_{21})$ 
\Comment{on the deviation path}

\vspace{3mm}

\Statex \textit{\textbf{Step 2: Iteration}}
\While{Price dynamics have not yet realigned with each other}
    \State $\theta_t$ is realized
    \State $(\Pi^*_1, \Pi^*_2) \gets (\Pi^*_1+\delta^{t}\pi^*_{1t}, \Pi^*_2+\delta^{t}\pi^*_{2t})$ 
    \State $(\Pi^D_1, \Pi^D_2) \gets (\Pi^D_1+\delta^{t}\pi^D_{1t}, \Pi^D_2+\delta^{t}\pi^D_{2t})$ 
    \State $t \gets t+1$
\EndWhile
\end{algorithmic}
\end{algorithm}

%%%%%%%%%%%%%%%%%%%%%%%%%%%%%%%%%%%%%
% \singlespacing
% \setlength\bibsep{0pt}
% \bibliographystyle{my-style}
% \bibliography{Placeholder}
\clearpage
\setcitestyle{numbers} % set the citation style to ``numbers''.
\bibliographystyle{chicago} % We choose the "plain" reference style
\bibliography{refs} % Entries are in the refs.bib file

\begin{thebibliography}{}
\bibitem[\protect\citeauthoryear{Abada, Harrington~Jr, Lambin, and Meylahn}{Abada et~al.}{2025}]{abada2025algorithmic}
Abada, I., J.~E. Harrington~Jr, X.~Lambin, and J.~M. Meylahn (2025).
\newblock Algorithmic collusion: where are we and where should we be going?
\newblock {\em Available at SSRN 4891033\/}.
\end{thebibliography}

\end{document}